\documentclass[aps,pra,preprint,showpacs,groupedaddress,superscriptaddress,longbibliography]{revtex4-1}

\usepackage[english]{babel}
\usepackage{float}
\usepackage{graphicx}
\usepackage[caption=false]{subfig}
\usepackage{amsmath}
\usepackage{amsfonts}
\usepackage{amssymb}
\usepackage{mathtools}
\usepackage{amsthm,enumitem}

\begin{document}

\title{Superexponential Interactions and the Dynamical Unfolding of Confined Degrees of Freedom}

 \author{Peter Schmelcher}
  \email{Peter.Schmelcher@physnet.uni-hamburg.de}
 \affiliation{Zentrum f\"ur Optische Quantentechnologien, Universit\"at Hamburg, Luruper Chaussee 149, 22761 Hamburg, Germany}
 \affiliation{The Hamburg Centre for Ultrafast Imaging, Universit\"at Hamburg, Luruper Chaussee 149, 22761 Hamburg, Germany}

\date{\today}

\begin{abstract}
We explore a two-body system with superexponential interactions that serves as a fundamental
building block for a route to complexity. While being of striking simplicity this highly nonlinear
interaction yields a plethora of intriguing properties and a rich dynamics. It exhibits a spatial region
where the dynamics occurs in a channel characterised by a transversally confined and longitudinally
unbounded motion and additionally two distinct regions where the dynamics is asymptotically free.
A deconfinement transition via two saddle points connects the dynamics in the channel with the asymptotically free motion.
The scattering functions show plateau and peak structures that can be interpreted in terms of
corresponding correlation diagrams. These are intimately related to the varying anharmonicity
of the transverse motion while moving along the longitudinal dimension of the channel.
We perform a comprehensive analysis of the scattering transition for energies below and above the
saddle points. Possible variants and extensions of the superexponential
interaction to many-body system are briefly discussed.
\end{abstract}

\maketitle

\section{Motivation and Introduction} 
\label{sec:introduction}

The route of describing nature by a bottom-up approach has been overwhelmingly successful
in physics. Elementary constituents and building blocks of matter and their interactions
are identified and used to explore the structure and dynamics of composite and more complex systems.
This holds for electrons and nuclei forming atoms \cite{Friedrich}, atoms binding
together and forming molecules \cite{Helgaker}, larger clusters \cite{Jellinek} or even nanostructures
\cite{Natelson} and crystals \cite{Ashcroft}. As a matter of fact, the fundamental forces between
such elementary constituents are typically of the appearance of an inverse power law such as
$\frac{1}{r^2}$ for Coulomb forces among charges and $\frac{1}{r^4}$ for dipolar forces among permanent
dipoles. An amazing and seemingly endless complexity of structures and properties emerges
from the interaction of many particles via these forces which is responsible for the variety
and diversity of phenomena which we observe in nature. 

A second well-known bottom-up route to complexity is the coupling of (non-)linear oscillators which
describe a plethora of phenomena. For coupled linear oscillators the dynamics is still integrable
and describes e.g. the multi-mode small amplitude vibrational dynamics of molecules \cite{Wilson}
or phonons in a bulk \cite{Ashcroft}. Coupled nonlinear and driven oscillators readily lead to a transition
from regularity to chaos characterized by a so-called mixed phase space with regular islands, chaotic
seas and fractal structures leading to stickiness and Levy flights \cite{Tabor,Strogatz,Reichl}.
This extends in (discrete) nonlinear models to localized excitations such as breathers \cite{Flach} and
even phononic frequency combs due to nonlinear resonances \cite{Cao}.

In view of the successful bottom up approach and encouraged by the corresponding emerging complexity,
one might ask the following question. Is there other interactions among corresponding fundamental building
blocks which would lead us via a different route to a different complex behaviour.
Here both the notion of fundamental building blocks and of their interactions
is not (necessarily) meant to be of microscopic origin, as it is the case for atoms and their compounds
describe above. They could be e.g. the result of an effective theory which incorporates a
hierarchy of degrees of freedom describing a highly nonlinear system.

In the above spirit, some preliminary steps have been taken very recently \cite{Schmelcher1,Schmelcher2}.
Firstly a driven power law oscillator \cite{Schmelcher1} has been explored where the exponent of the oscillator potential
is harmonically oscillating in time according to $V(q) \propto |q|^{\beta(t)}$ with $ \beta(t) \propto \sin \rm{\omega} t$
thereby covering a broad spectrum of anharmonicities during the cyclic
evolution. This oscillator leads to a two-component phase space. Bounded motion occurs with
an underlying mixed regular chaotic phase space and its characteristic dynamical consequences.
The second component in phase space corresponds to an unbounded motion which exhibits an
exponential net growth of the corresponding energies leading to a tunable exponential acceleration.
In a second step the superexponential self-interacting (SSO) oscillator \cite{Schmelcher2} has been introduced and analyzed. 
For this oscillator both the base and the exponent depend on the dynamical variable (coordinate)
of the oscillator leading to the potential $V(q) \propto |q|^q$. In this case it is left to the intrinsic
time-evolution of the dynamical oscillator which instantaneous potential it 'feels'.
Opposite to standard oscillators such as the (an-)harmonic confining oscillator with
$V(q) \propto q^{2n}$ with $n \in \mathbb{N}$ the SSO combines both scattering and confined
periodic motion with an exponentially varying nonlinearity. Specifically the SSO potential exhibits
a transition point with a hierarchy of singularities of logarithmic and power law character
leaving their fingerprints in the agglomeration of its phase space curves. The period of the
SSO consequently undergoes a crossover from decreasing linear to a nonlinearly increasing
behaviour when passing the transition energy. 

In the present work we perform a significant step forward in the above discussed route to complexity.
As a key ingredient and building block we introduce a superexponential interaction
potential (SEP) ${\cal{V}}(q_i,q_j) = |q_i|^{q_j}$ for the degrees of freedom $q_i$
and $q_j$. Obviously such an interaction does not respect common symmetries for the fundamental forces in nature
such as the translational or permutational invariance. Therefore the degrees of freedom $q_i$
would typically belong to a more complex subunit which we still call in the following (effective) particles.
We explore and analyze the two-body problem as a key ingredient for all further
investigations on larger systems. The extreme spatially varying nonlinearity of the
SEP which is experienced by the particles via their intrinsic dynamics leads to a very rich and
uncommon behaviour. The SEP possesses a channel of hybrid confined and unbounded motion with
spatially varying anharmonicities which is separated via two saddle points from two regions of unbounded free motion
which are separated by a repulsive barrier. We explore the dynamics on the level of single trajectories
and ensemble properties below and above the saddle point energies thereby developing a detailed
understanding of the deconfinement transition from confined channel dynamics to the asymptotically
free motion. This opens up the perspective of considering the SEP and the many possible modifications
of it as a fundamental building block of a dynamical complex network.

In detail we proceed as follows. In section \ref{sec:hamiltonian} we describe our setup and analyze the properties
of the superexponential two-body potential. This section is central to understand the major
differences of the SEP as compared to many standard two-body potentials. Section \ref{sec:dynamics} contains
an elaborate discussion of the dynamics taking place in the SEP. Subsection \ref{sec:dynamicsa} is dedicated to
a trajectory-based scattering analysis in the hybrid confining channel below the energy
of the saddle points. Subsection \ref{sec:dynamicsb} analyzes the ensemble behaviour in this channel and the most striking
properties due to the spatially and dynamically varying anharmonicities. The scattering dynamics for
energies above the saddle points leading to a deconfinement transition is investigated in subsection \ref{sec:dynamicsc}.
Finally, section \ref{sec:conout} contains our conclusions and outlook briefly addressing the many perspectives
which open up on basis of our two-body investigation. This includes not only the extension to 
the many-body problem but also the rich possible variations of its two-body ingredient based on the SEP
touching upon higher dimensions, varying exponents and geometries.

\section{The Superexponential Two-Body Potential} \label{sec:hamiltonian}

\begin{figure}
\hspace*{-1cm} \parbox{10cm}{\includegraphics[width=10cm,height=7cm]{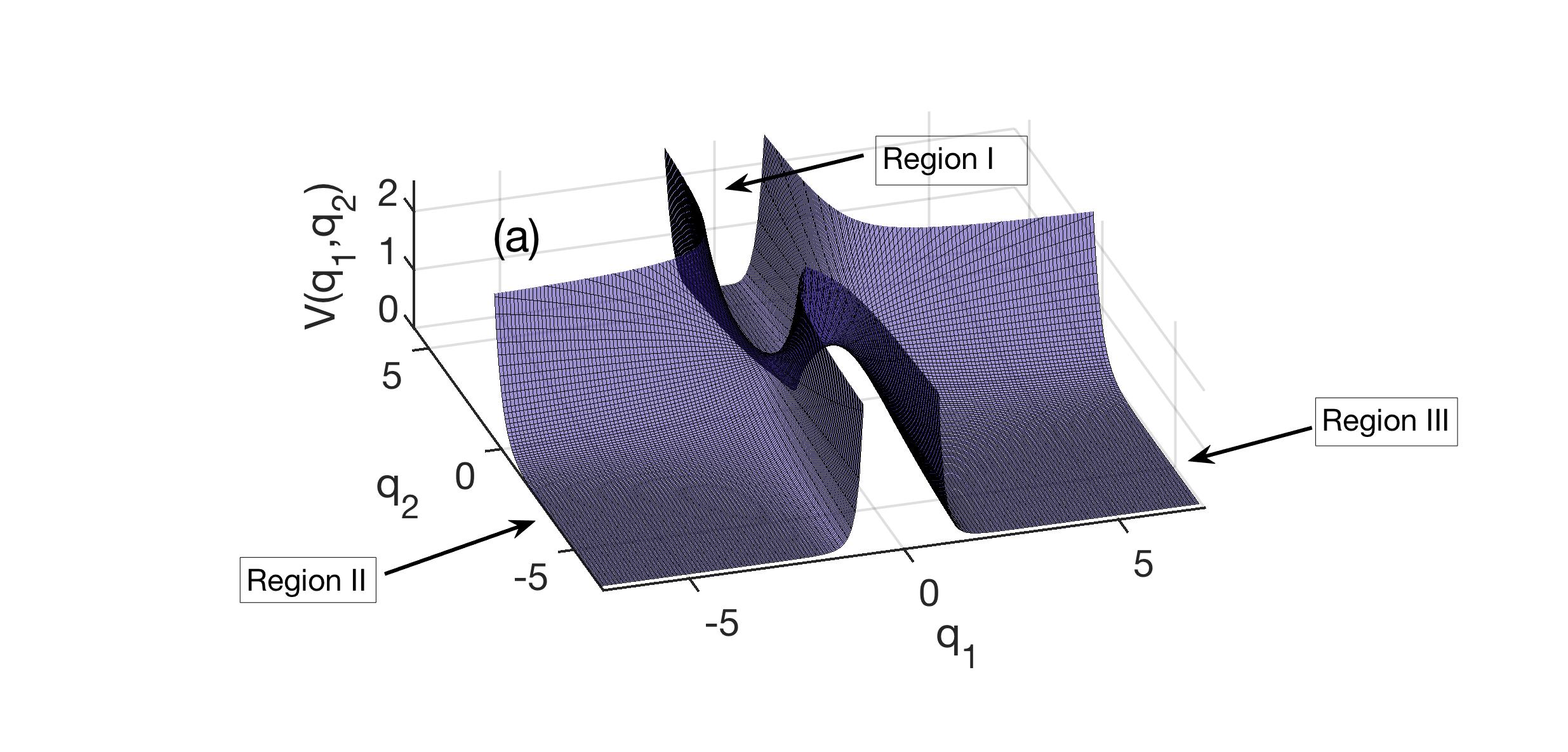} } 
\hspace*{-0.7cm} \parbox{6cm}{\vspace*{1cm} \includegraphics[width=12cm,height=7cm]{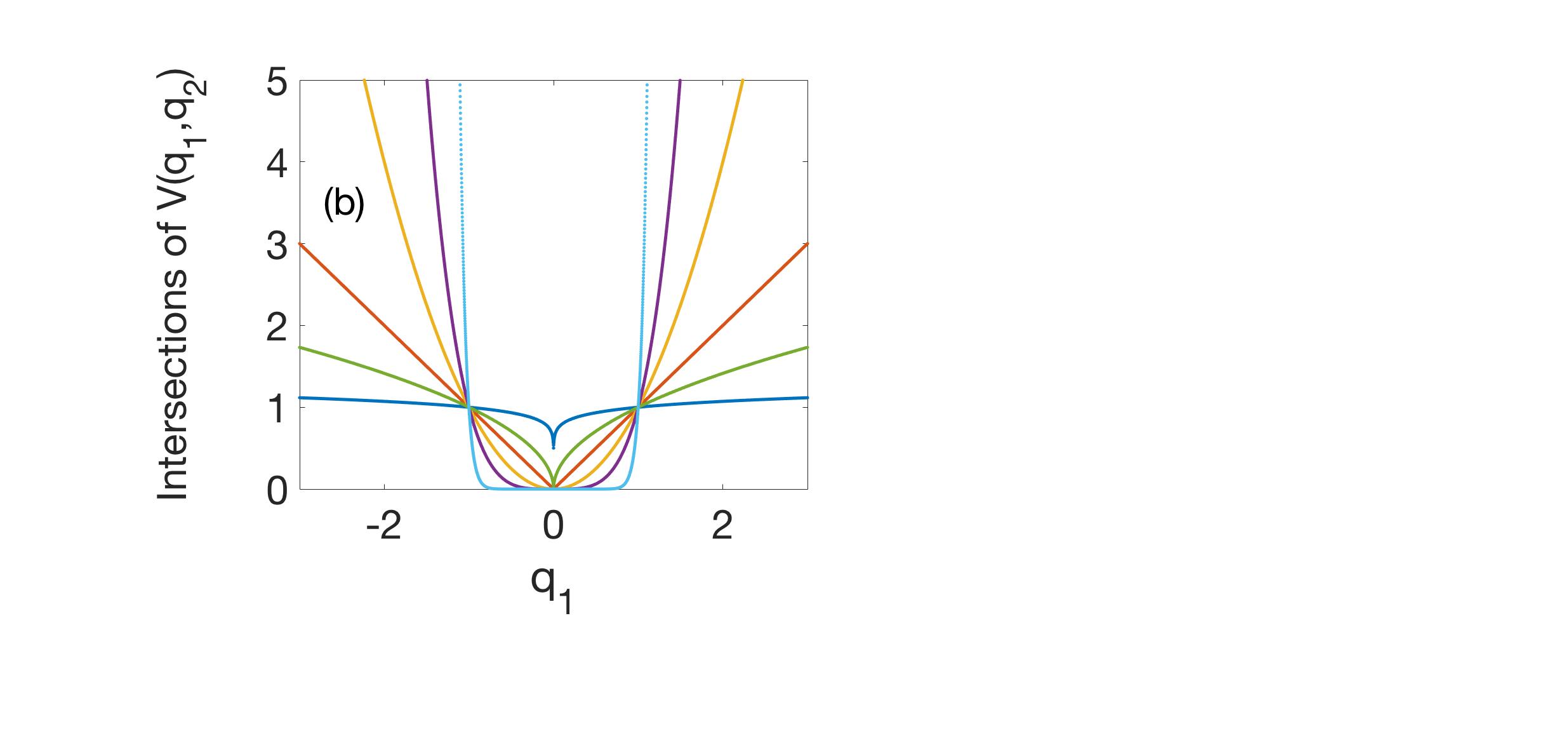}} 
\parbox{6cm}{\vspace*{-1cm} \includegraphics[width=11cm,height=6cm]{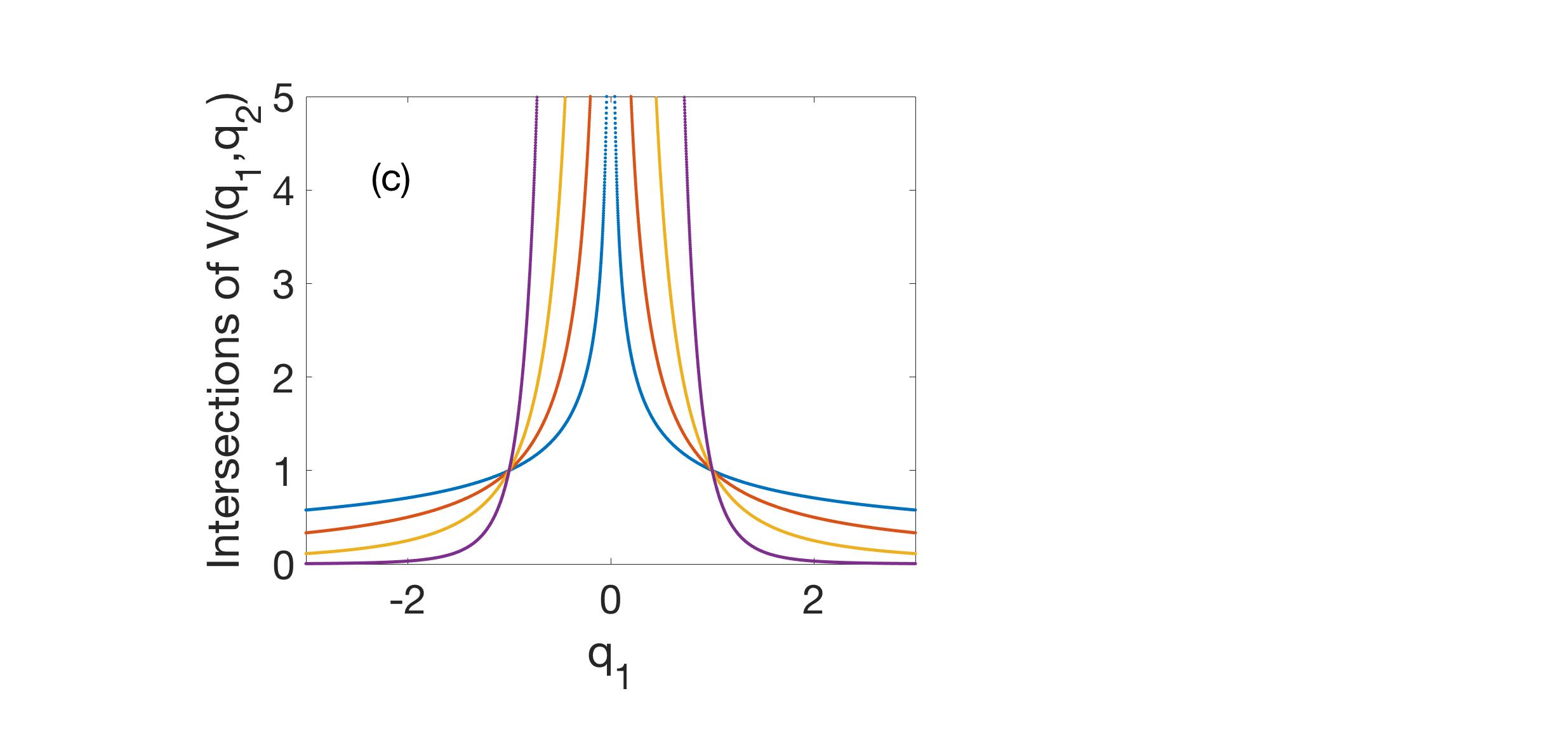}} 
\caption{(a) The potential energy surface ${\cal{V}}(q_1,q_2)$. (b) Intersections
of the potential energy energy surface ${\cal{V}}(q_1,q_2)$ along the $q_1$ coordinate 
for fixed values of $q_2$. Curves from bottom to top in the interval of $-1< q_1 < 1$
correspond to the values $q_2 = 16,4,2,1,0.5,0.1$. (c) same as in (b) but for the 
values $q_2=-0.5,-1,-2,-5$.}
\label{fig1}
\end{figure}

In this section we introduce the Hamiltonian of our two-body problem with superexponential interactions
and discuss its major properties. This will serve as a basis for the forthcoming investigations on the
dynamics in section \ref{sec:dynamics}. The Hamiltonian of our two-body problem reads as follows

\begin{equation}
{\cal{H}} = \frac{p_1^2}{2} + \frac{p_2^2}{2} + |q_1|^{q_2}
\label{eq:hamiltonian}
\end{equation}

the coordinates $q_1,q_2$ and their kinetic momenta $p_1,p_2$ should be considered as
belonging to two effective particles or, in general, effective degrees of freedom describing the motion of the
fundamental building blocks of our two-body system. The last term of the above Hamiltonian (\ref{eq:hamiltonian})
constitutes the superexponential potential ${\cal{V}} = |q_1|^{q_2}$ whose properties are of central
importance to all what follows. In contrast to many well-established fundamental interaction potentials,
such as the power law potentials $\propto q^n$ with $n \in \mathbb{Z}$, the SEP possesses a peculiar shape
in the sense that both the base and the exponent of it depend on two different dynamical degrees of freedom
$q_1$ and $q_2$ respectively. This automatically implies that any dynamics taking place in the coordinate
$q_2$ leads to a different potential instantaneously experienced by the degree of freedom $q_1$ and vice versa. 
As a result of this superexponential coupling the instantaneous exponent can vary from extremely confining
($q_2 \rightarrow \infty$) to extremely repelling ($q_2 \rightarrow - \infty$). Figure \ref{fig1} shows 
a potential energy surface plot (Figure \ref{fig1}(a)) of ${\cal{V}}$ together with corresponding
intersections of it along the $q_1$ coordinate (Figure \ref{fig1}(b,c)) for various fixed values of 
the coordinate $q_2$. 

Focusing on the surface plot of ${\cal{V}}$ one observes that there are three different regions separated
by potential barriers. Region I occurs for positive values of $q_2$ meaning that the SEP has a positive
exponent and the motion is confined along the $q_1$ coordinate. This leads to a channel of the potential clearly
visible in Figure \ref{fig1}(a) which extends from $q_2=0$ to $q_2 = + \infty$. Consequently the motion
inside this confining channel (CC) combines a bounded confined motion of $q_1$ with an unbounded scattering motion along $q_2$,
both of them being superexponentially coupled. The latter leads to the fact that the CC is extremely
inhomogeneous as can be seen in Figure \ref{fig1}(b) focusing on the intersection curves for positive values
of $q_2$. For large positive values of $q_2$ the CC possesses a hard wall box-like confinement which it
reaches asymptotically for $q_2 \rightarrow + \infty$ and for which case the $q_2$ motion is a free
ballistic motion decoupled from the $q_1$ oscillatory box motion. Decreasing the value of $q_2$ leads to a decrease
of the anharmonicity covering continously all values of the exponent passing on to a harmonic confinement
$q_2=2$, eventually to a linear one $q_2=1$. At $q_2=1$ the second derivative $\frac{\partial^2 {\cal{V}}}{\partial
q_2^2}$ turns from overall positive to negative with further decreasing values of $q_2$.
Consequently for $q_2 < 1$ the CC develops a kink at $q_1=0$ and flattens out towards a constant
potential which it reaches for $q_2=0$. Still, the CC maintains an increasingly narrow part around $q_1=0$
of significant depth whose width goes to zero for $q_2 \rightarrow 0$.

For $q_2 < 0$ ${\cal{V}}$ exhibits two other distinct regions II and III (see Fig. \ref{fig1} (a)). They are separated by 
a potential barrier with a (singular) maximum at $q_1=0$. This barrier is naturally of purely
repulsive character and widens with a decreasing value of $q_2$, while becoming an infinite repulsive square barrier
on the interval $q_1 \in [-1,+1]$ along the degree of freedom $q_1$ for $q_2 \rightarrow - \infty$
(see Figure \ref{fig1}(c)).
In both regions II and III the SEP flattens out
with increasing values of $|q_1|$ and a given negative value of $q_2$. This implies that the motion in regions
II and III becomes asymptotically free. The difference between the two regions is the fact that
region II implies that both particles move in a correlated manner to the same direction ($q_1,q_2 < 0$)
whereas in region III they move to opposite directions. Region I with the CC motion and regions II and III
with an unconfined asympotically free motion are separated by two saddle points of the SEP. Taking the partial
derivatives of ${\cal{V}}$ and demanding them to be zero yields $q_1= \pm 1$ and $q_2=0$ for the positions of
the extrema. The determinant of the corresponding second derivatives is negative thereby providing us with the
saddle point character of these extrema. The energies of the saddle points are $E=1$ and for energies below
this energy the motion in the regions I,II,III is disconnected, i.e. an incoming trajectory in the CC would
be exclusively reflected (see section \ref{sec:dynamics} for the investigation of the corresponding dynamics) and
not transmitted to the regions II and III with their asymptotically free behaviour. For energies above the
saddle point energies transmission is possible and the regions I and II,III are dynamically connected.

In essence, while the appearance of the SEP is strikingly simple it shows an unexpectedly rich geometrical structure
with a transition from confined channel motion via saddle points to two distinct regions of asympotically
free motion within which the particles carry different correlations. The intricate highly nonlinear coupling
of the degrees of freedom occuring in the SEP is also manifest when formulating it in center of mass and
relative coordinates, which are given by $Q=\frac{q_1+q_2}{2}$ and $q=q_1-q_2$ respectively. The SEP then
reads ${\cal{V}}= |Q+\frac{q}{2}|^{Q-\frac{q}{2}}$ which demonstrates the intricate coupling and nonseparability
of the center of mass and internal motion of the SEP. Therefore, a treatment in these coordinates does
not offer any advantage.

Some notes are adequate at this place. The SEP is neither short-ranged nor long-ranged according to the traditional
classification of interaction potentials. It intricately couples the degrees of freedom in a superexponential way
thereby enabling the above-discussed variety of behaviour within a simple two-body systems. It combines
confining channel and free boundary conditions within a single interaction term. These aspects render
the SEP a very promising candidate for a complex geometrical network of corresponding many-body systems as we
shall discuss later on in some more detail. The route to complexity is here obviously very much different from
some known routes mentioned previously. As a final note we mention that the recently introduced and explored
superexponential self-interacting oscillator \cite{Schmelcher2} is a specialization of the present two-body 
and SEP based problem. Degenerating the two degrees of freedom $q_1,q_2$ in the sense of the choice 
$q_1=q_2=q$ leads to the one-dimensional interaction potential ${\cal{V}}(q,q)=|q|^q$ whose peculiar
properties have been explored in ref.\cite{Schmelcher2}. We remark that this potential corresponds
to the intersection of our SEP potential ${\cal{V}}(q_1,q_2)$ along its diagonal which does not pass through its
saddle points.

\section{Two-Body Dynamics for Superexponential Interactions} \label{sec:dynamics}

This section is dedicated to the investigation of the dynamics in the SEP. In subsection \ref{sec:dynamicsa}
we will first analyze the features of individual trajectories. Our focus is on incoming scattering trajectories
in region I i.e. from the asymptotics of the confined channel and for energies below the saddle point energies.
In subsection \ref{sec:dynamicsb} the corresponding trajectory ensemble properties are investigated and interpreted.
Finally, for energies above the saddle point energies we explore the scattering dynamics of both individual trajectories
and in particular of ensembles in the subsection \ref{sec:dynamicsc}: this dynamics connects the channel region I
to the asymptotically free motion in the regions II and III.

\begin{figure}
\hspace*{-3.4cm} \parbox{10cm}{\includegraphics[width=13cm,height=7cm]{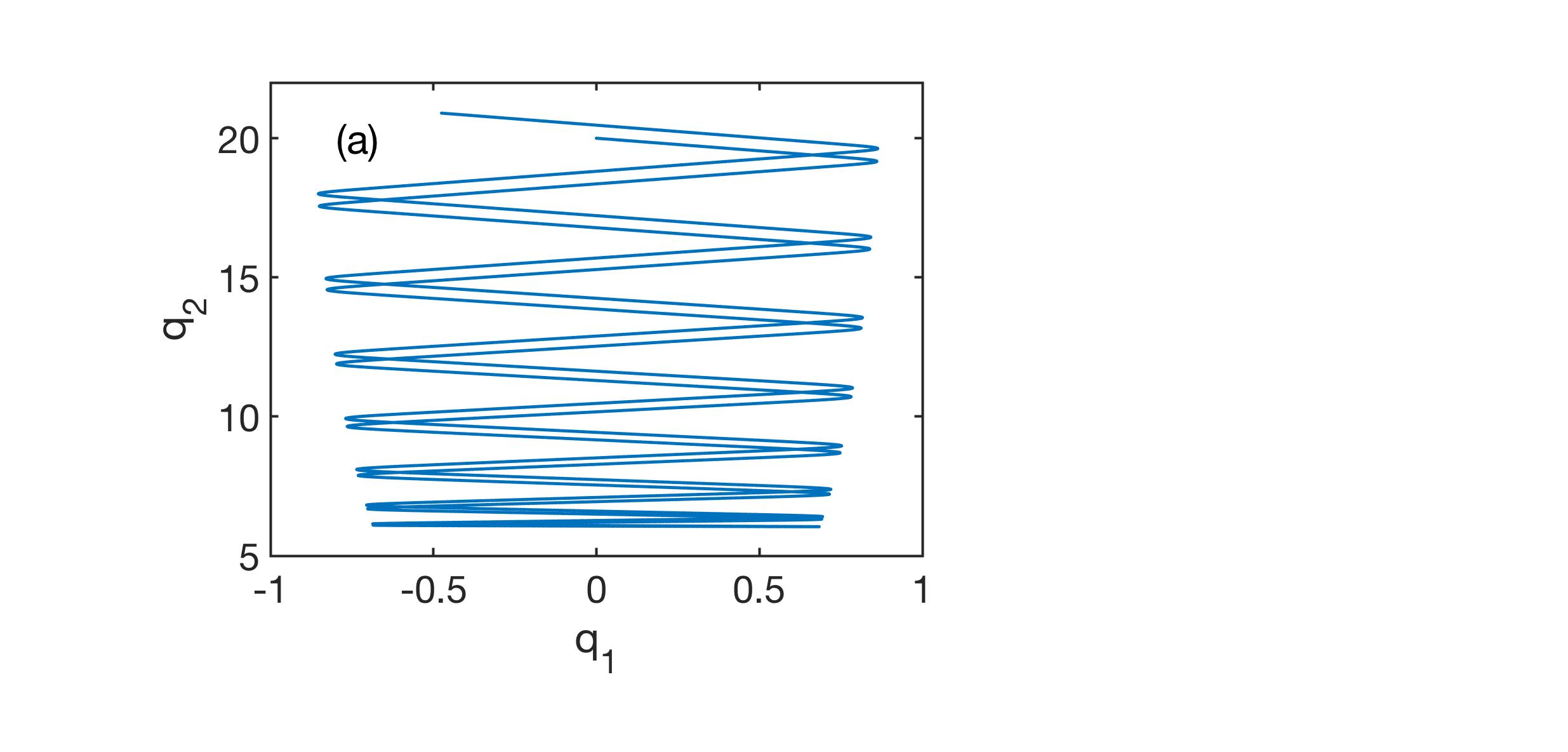} } 
\hspace*{-2.2cm} \parbox{6cm}{\includegraphics[width=13cm,height=7cm]{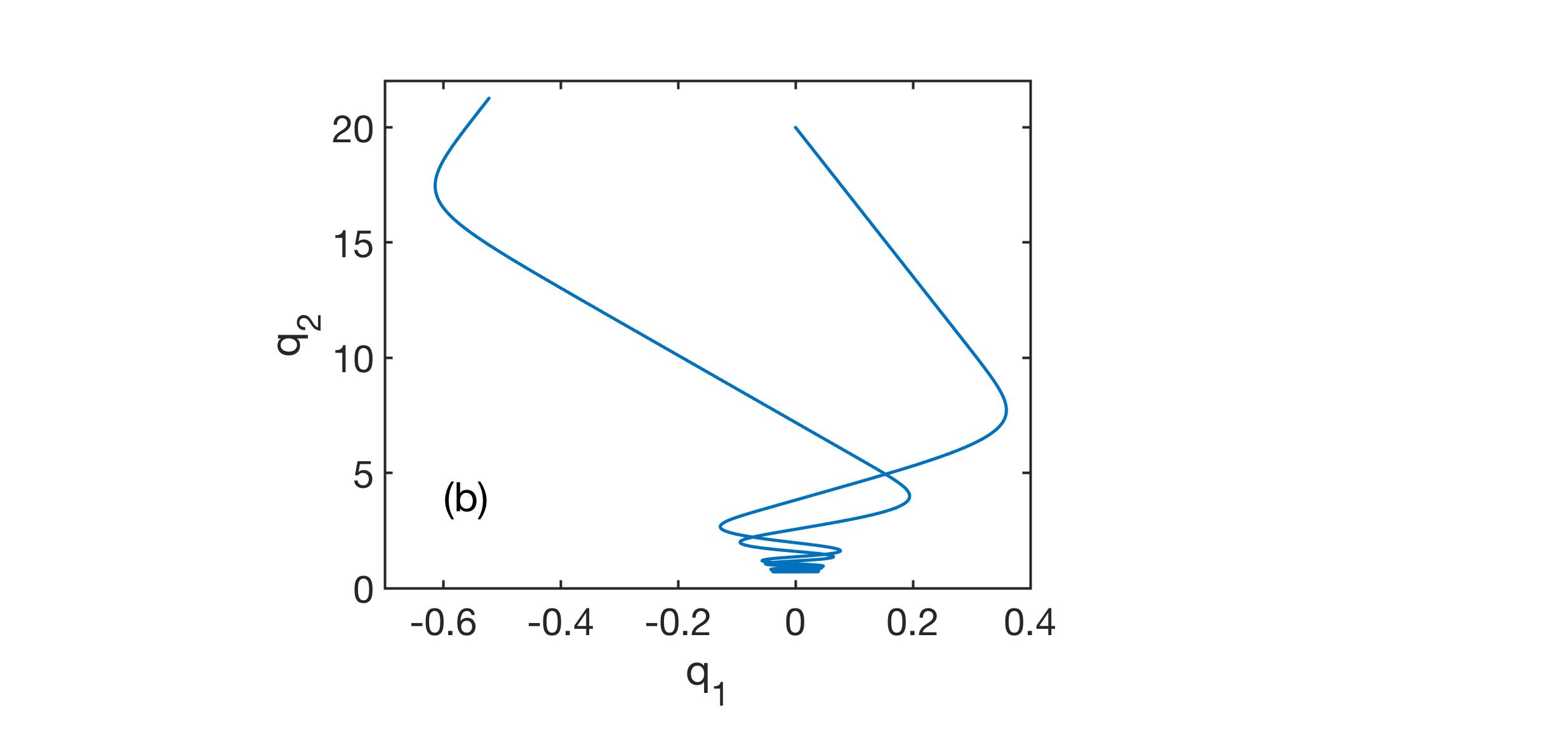}} \\
\hspace*{-2cm} \parbox{10cm}{\includegraphics[width=11cm,height=7cm]{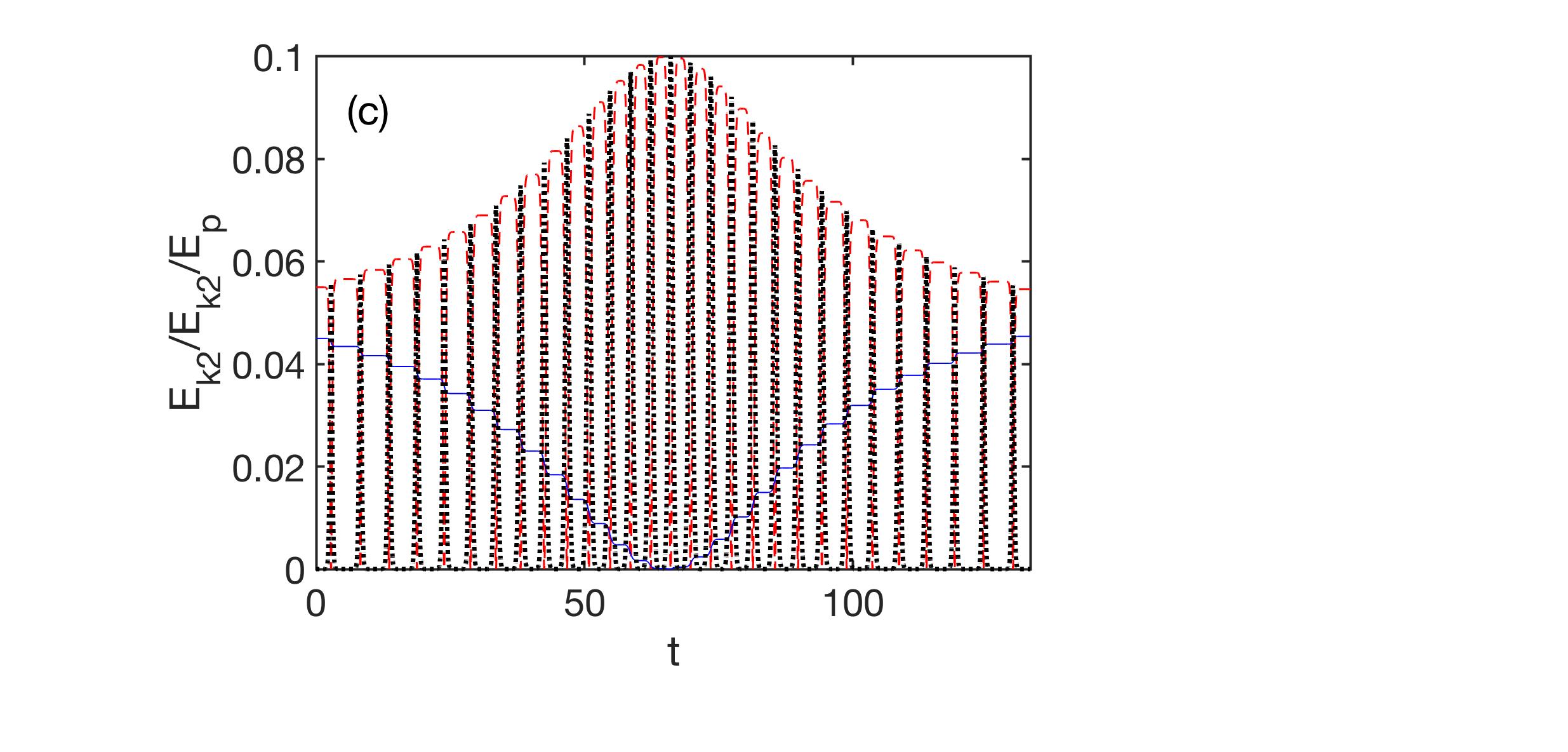} } 
\hspace*{-1.7cm} \parbox{6cm}{\includegraphics[width=13cm,height=7.7cm]{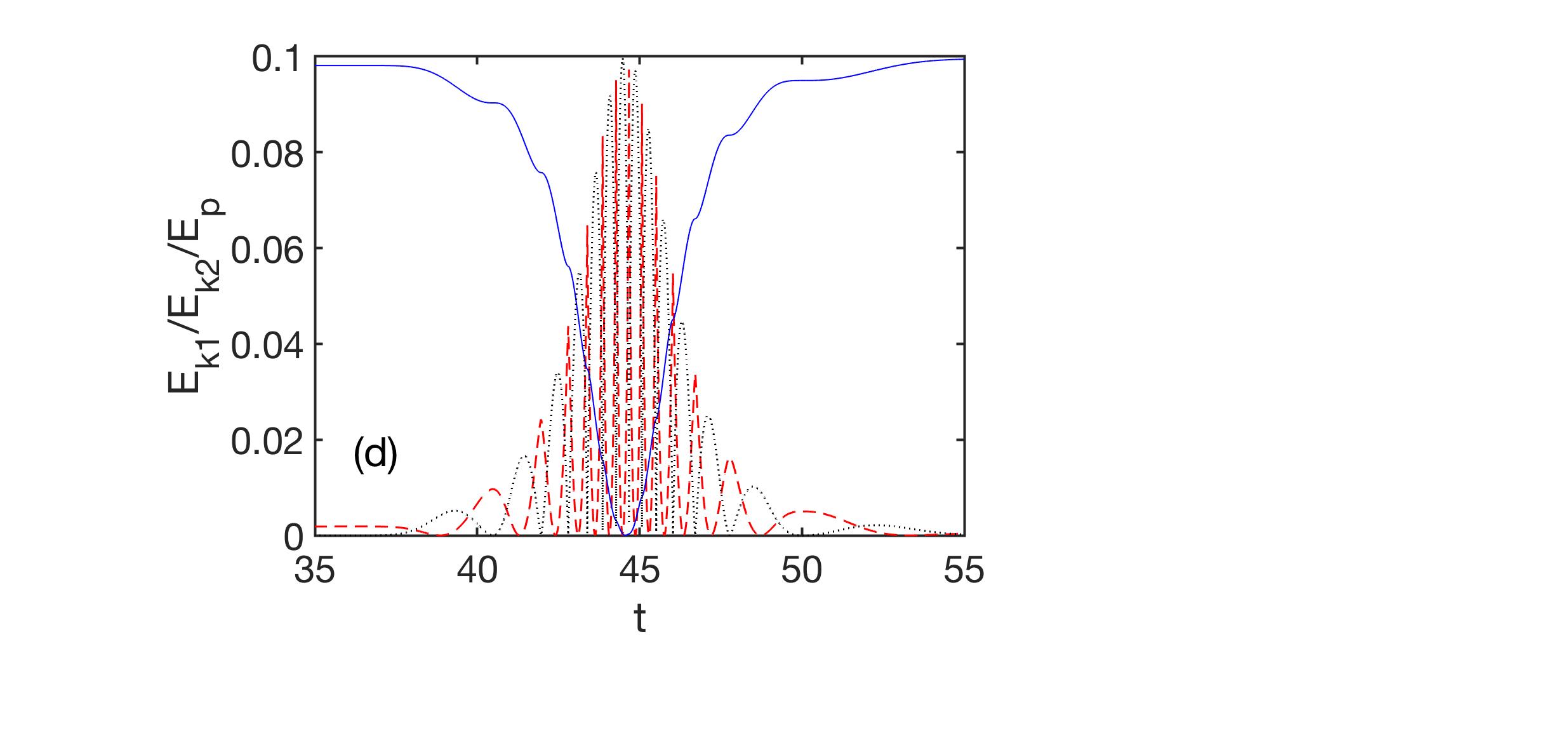}} 
\caption{(a) Single trajectory in ($q_1,q_2$) plane for the initial conditions (incoming
channel region I) $q_1=0, q_2=20, p_2=-0.3$ for the energy $E=0.1$. The corresponding
time evolution of the kinetic energy $E_{k1}=\frac{p_1^2}{2}$ (red dashed line) and
of $E_{k2}=\frac{p_2^2}{2}$ (blue solid line) as well as the potential energy 
$E_p=|q_1|^{q_2}$ (black dotted line) is shown in subfigure (c). Subfigure (b)
shows a corresponding trajectory for the same initial conditions except $p_2=-0.447$
which is strongly ($q_2$) forward directed. Subfigure (d) shows the corresponding
kinetic and potential energies as in subfigure (c).}
\label{fig2}
\end{figure}

\subsection{Dynamics in the Confined Channel: Individual Trajectories} \label{sec:dynamicsa}

The dynamics is described by the Hamiltonian equations of motion belonging to the Hamiltonian
(\ref{eq:hamiltonian}). To regularize these equations of motion which possess a singularity
for $q_2 < 0$ at $q_1 = 0$, we introduce a regularization parameter $\epsilon > 0$ for the SEP
which now reads ${\cal{V}} (q_1,q_2;\epsilon) = (\sqrt{q_1^2 + \epsilon})^{q_2}$ and facilitates
the numerical integration. This yields the following equations of motion

\begin{eqnarray}
\dot{p}_1 = - \frac{q_1 q_2}{(q_1^2 + \epsilon)} \left( \sqrt{q_1^2 + \epsilon} \right)^{q_2}\\
\dot{p}_2 = - \left( \sqrt{q_1^2 + \epsilon} \right)^{q_2} \rm{ln} \left(\sqrt{q_1^2 + \epsilon} \right)
\end{eqnarray}

Typical values for the regularization parameter are $\epsilon = 10^{-8}$. Our numerical results both
on the individual trajectory level as well as the ensemble behaviour are independent on this regularization
parameter.

We focus in this subsection on incoming ($p_2 < 0$) trajectories in the CC of region I for energies below
the saddle point energies. These trajectories are all back reflected within the CC and escape asymptotically
to $q_2 \rightarrow - \infty$. As indicated above (see section \ref{sec:hamiltonian}), the trajectories 
are initialized for large values of $q_2$ such that they propagate during the initial phase of their
dynamics in an approximately box-like strongly anharmonic channel. In the course of the scattering dynamics the
transverse channel confinement continuously changes its exponent ultimately covering all anharmonicities,
the close to harmonic case and the linear case while finally developing the above-discussed (see
section \ref{sec:hamiltonian}) case of $q_2 < 1$ with a flattening and narrowing well around $q_1 = 0$.
It depends on the initial conditions, more
specifically on the ratio of the longitudinal ($p_2$) to transverse ($p_1$) momenta, and on the total
energy $E$ what the range of exponents is that is experienced by a specific trajectory. 
We will focus here on two major examples of individual trajectories, which are the cases of
(i) momenta $p_1$ and $p_2$ which are comparable in magnitude and (ii) a forward direction scattering
process for which $|p_2|>>|p_1|$. Figures \ref{fig2}(a,b,c,d) show the corresponding motions in the
($q_1,q_2$) plane together with the time evolution of the kinetic energies $E_{k1} = \frac{p_1^2}{2}$,
$E_{k2} = \frac{p_2^2}{2}$ and the potential energy $E_{p} = |q_1|^{q_2}$.

Figure \ref{fig2}(a) shows a typical trajectory in the ($q_1,q_2$) plane originating from $q_2=20$ with
a momentum $p_2 < 0$ according to the above case (i) for a comparatively low energy $E=0.1$.
Starting at the outer parts of the confining 
channel where the transversal anharmonicity is extremely pronounced and a box-like confinement is
present it travels inward (towards $q_2 = 0$) while performing transversal oscillations whose
amplitudes decrease with decreasing value of $q_2$. At $q_2 \approx 6$
it is backreflected in the channel and travels outward. Figure \ref{fig2}(c) shows the 
time evolution of the corresponding kinetic $E_{k1},E_{k2}$ and potential $E_{p}$ energies.
Overall the kinetic energy $E_{k1}$ increases and $E_{k2}$ decreases for the incoming channel trajectory
and they exhibit a maximum/minimum at the closest 'collision' point when the trajectory is reflected back.
Subsequently the reverse process happens. On top of this envelope behaviour $E_{k1}(t)$ and $E_{k2}(t)$
show a sequence of plateaus with rapid changes in between them. The widths of these plateaus decreases
with decreasing value of $q_2$ and their shape turns into a smooth peak structure. The potential energy
$E_p(t)$ mediates the energy transfer between the kinetic energies $E_{k1}$ and $E_{k2}$ and shows
pronounced sharp peaks for the times when the transition between the plateaus happens. The plateau structure
stems from the traversing of the trajectory of the bottom inner part of the channel during a transversal $q_1$
oscillation. In this channel part the potential energy contribution is very small and subsequently 
each of the kinetic energies $E_{k1},E_{k2}$ is approximately conserved. 

Figure \ref{fig2}(b) shows a trajectory again in the ($q_1,q_2$) plane emanating in the outer parts
of the confined channel but now for the case (ii) of a dominant momentum $p_2$. This corresponds to a 
a very much forward (towards the scattering center around ($q_1=q_2=0$) directed scattering process. As a
consequence the trajectory can now enter much deeper into the confining channel in the sense that
it experiences the narrowing of the channel for much smaller values of the exponents $q_2$. Indeed,
Figure \ref{fig2}(b) shows the squeezing of the dynamics impressively i.e. the systematic suppression
of the amplitude of the transversal oscillations with decreasing value of $q_2$ which is the longitudinal
channel coordinate. After only a single transversal oscillation the squeezing takes over.
The minimal value of $q_2$ corresponding to the turning point of the motion is now approximately one. 
As a consequence it can be observed in Figure \ref{fig2}(d) that the plateau structure of the
kinetic energies $E_{k1},E_{k2}$ in case (i) (Figure \ref{fig2}(c)) is no more present in this case 
but a localized sequence of peaks with strongly varying maximal values in the course of the time
evolution. Concerning the flow of energy between the kinetic and potential energies similar statements
like the above-ones hold. It is important to note that for both above cases the incoming and outgoing
kinetic energies $E_{k1},E_{k2}$ are approximately equal.

\begin{figure}
\hspace*{-3.8cm} \parbox{10cm}{\includegraphics[width=13cm,height=7.6cm]{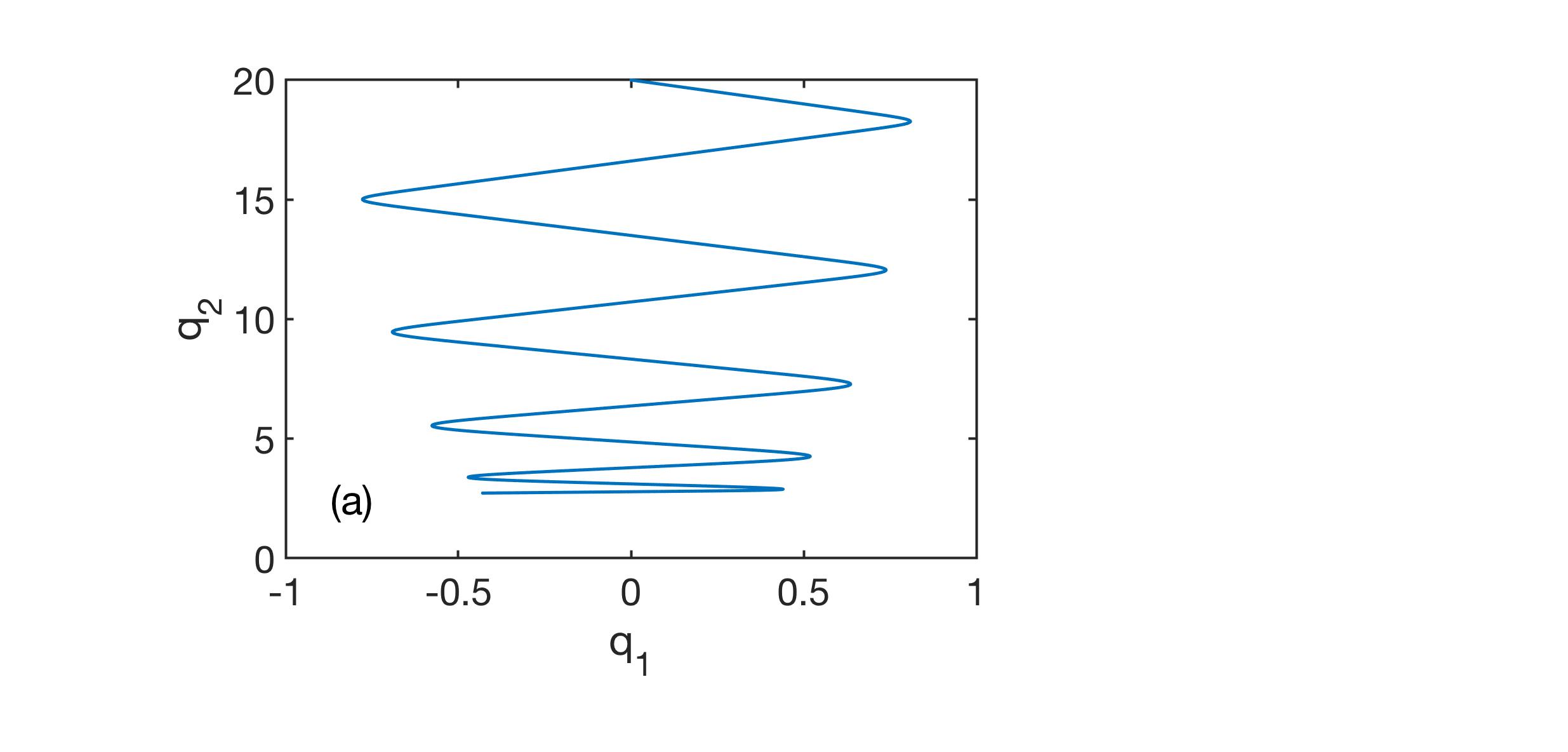}}
\hspace*{-1.6cm} \vspace*{-0.8cm} \parbox{6cm}{\vspace*{0.4cm} \includegraphics[width=14cm,height=8.0cm]{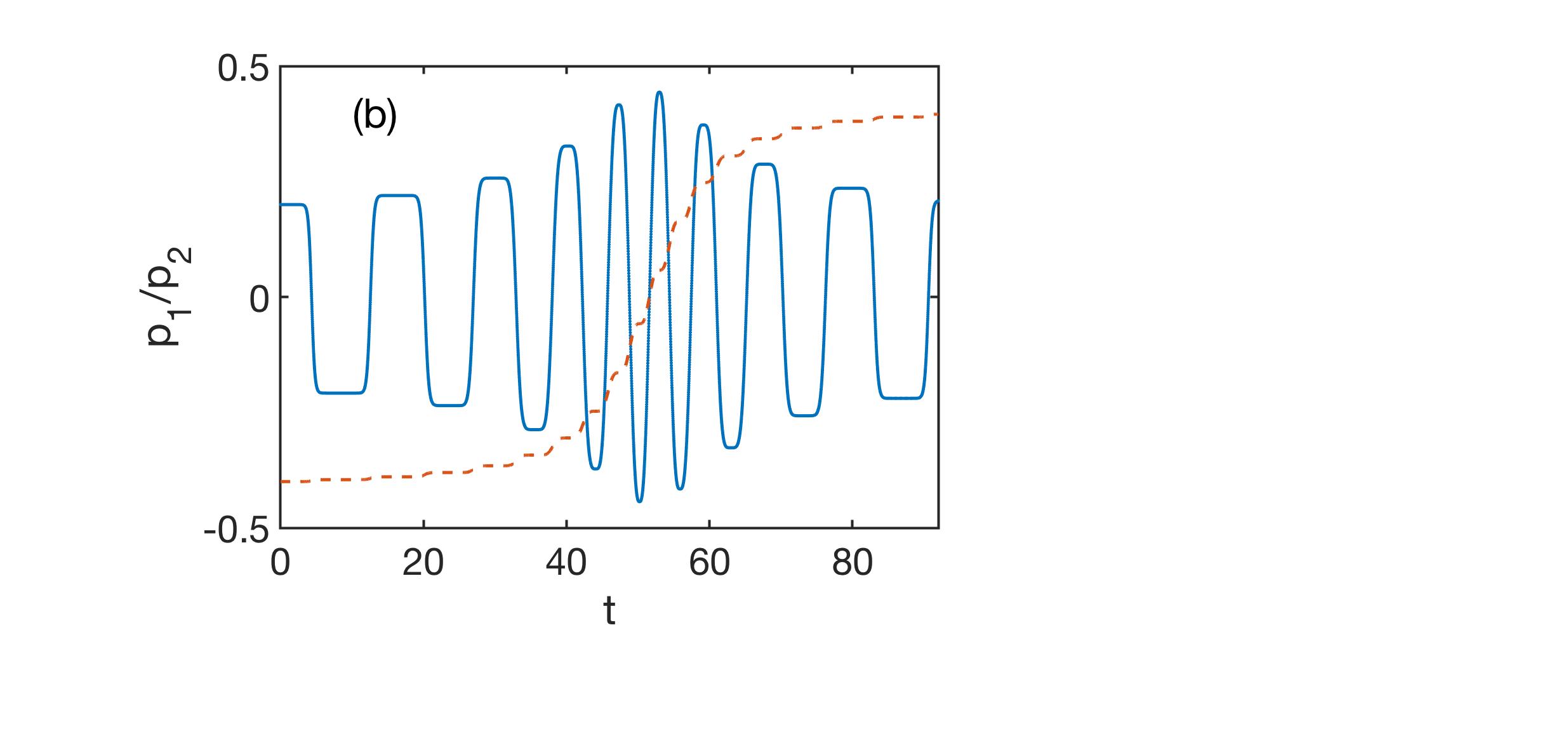}}
\caption{(a) Single return trajectory in ($q_1,q_2$) plane for the initial conditions (incoming
channel region I) $q_1=0, q_2=20, p_2=-0.4$ for the energy $E=0.1$. (b) The corresponding
time evolution of the momenta $p_1$ (blue solid line) and $p_2$ (red dashed line).}
\label{fig3}
\end{figure}

Inspecting incoming trajectories from the (asymptotic) box-like channel one realizes that there is a certain
class of trajectories which are reflected back onto themselves in configuration space ($q_1,q_2$), which we
call  return trajectories. Such a case is shown in Figure \ref{fig3}(a) with the same initial conditions
as in Figure \ref{fig2}
but for the incoming momentum value $p_2=-0.4$. Incoming and outgoing channel motion in the ($q_1,q_2$)
plane equal to a very good approximation. Let us analyze this situation in some detail. The turning or
closest collision point of a channel trajectory corresponds to the minimum value of $q_2$ for which
$\dot{q}_2(t_0)=0$. Asking for an exact return trajectory requires to reverse the motion at $t_0$ which
can be shown to yield the condition $\dot{q}_1(t_0)=\dot{q}_2(t_0)=0$ for the two degrees of freedom
$q_1$ and $q_2$. Figure \ref{fig3}(b) demonstrates this for the case of the trajectory shown in Figure
\ref{fig3}(a) in the ($q_1,q_2$) plane. To a very good approximation for this case both velocities
$\dot{q}_1=\dot{q}_2$ becomes zero at the same time $t_0 \approx 52$. The occurrence of such (approximate)
return trajectories is supported by the strong oscillatory character of the transversal channel motion
$q_1(t)$ rendering it possible to meet the above condition at the turning point of the $q_2$-motion.
Since the point of return reflection implies the vanishing of both kinetic energies $E_{k1}(t_0),E_{k2}(t_0)$
the total energy is equal to the potential energy $E=E_p(t_0)$. The latter means that the location
of this return reflection takes place at the outer parts of the channel where the potential energy
becomes signficant (see also Figure \ref{fig3}(a)). A channel trajectory that is strongly squeezed
(see Figure \ref{fig2}(b)) which enters deep into the region of the narrowed channel is much more 
sensitive to the detailed initial conditions and consequently the achievement of a return trajectory
becomes a fine tuning process. 

\begin{figure}
\hspace*{-3.8cm} \parbox{10cm}{\includegraphics[width=13cm,height=7.6cm]{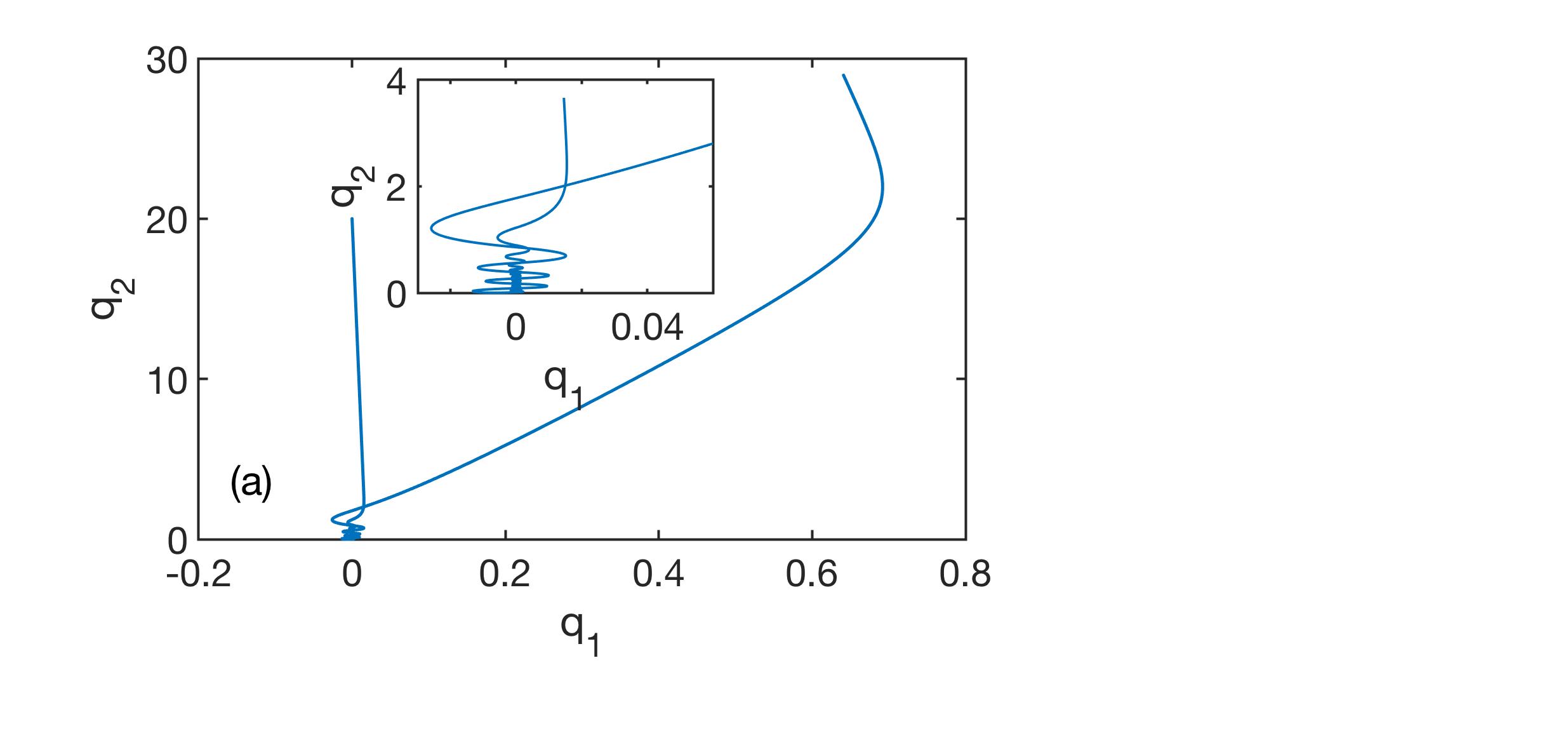}}
\hspace*{-1.6cm} \vspace*{-0.8cm} \parbox{6cm}{\vspace*{0.4cm} \includegraphics[width=14cm,height=8.0cm]{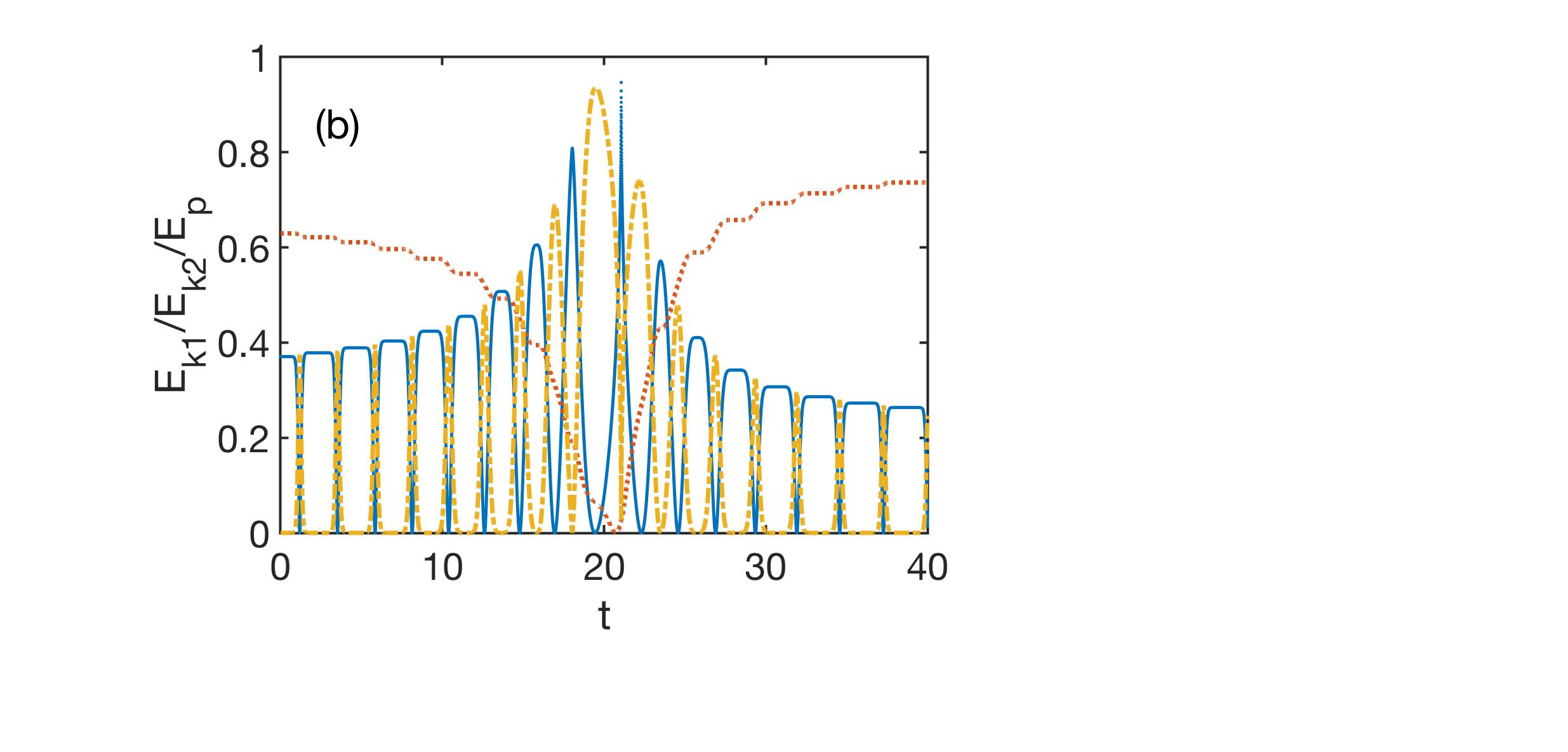}}
\caption{(a) A squeezing trajectory in the ($q_1,q_2$) plane for the initial conditions (incoming
channel region I) $q_1=0, q_2=20, p_2=-1.414213$ for the energy $E=1$. Inset: Magnification
around the turning point. (b) Time evolution of the kinetic energies $E_{k1}$ (blue solid line),
$E_{k2}$ (red dotted line) and the potential energy $E_p$ (orange dashed dotted line) for $p_2=-1.122$.}
\label{fig4}
\end{figure}

Let us now focus on large energies of the scattering processes in the channel region I. Figure \ref{fig4}(a)
shows a strongly squeezed trajectory for $E=1$ which is forward focused ($p_2=-1.414213$). The strong 
asymmetry for the incoming and outgoing motion, in particular also close to the turning point (see
inset of Figure \ref{fig4}(a)), is clearly visible. Figure \ref{fig4}(b) shows the time evolution
of a trajectory with $p_2=-1.122$ for otherwise identical initial conditions and total energy. Here
a strong asymmetry with respect to the asymptotic incoming and outgoing kinetic energies can be observed
which indicates the inelasticity of the underlying process.

\begin{figure}
\hspace*{-3.8cm} \parbox{10cm}{\includegraphics[width=13cm,height=7.6cm]{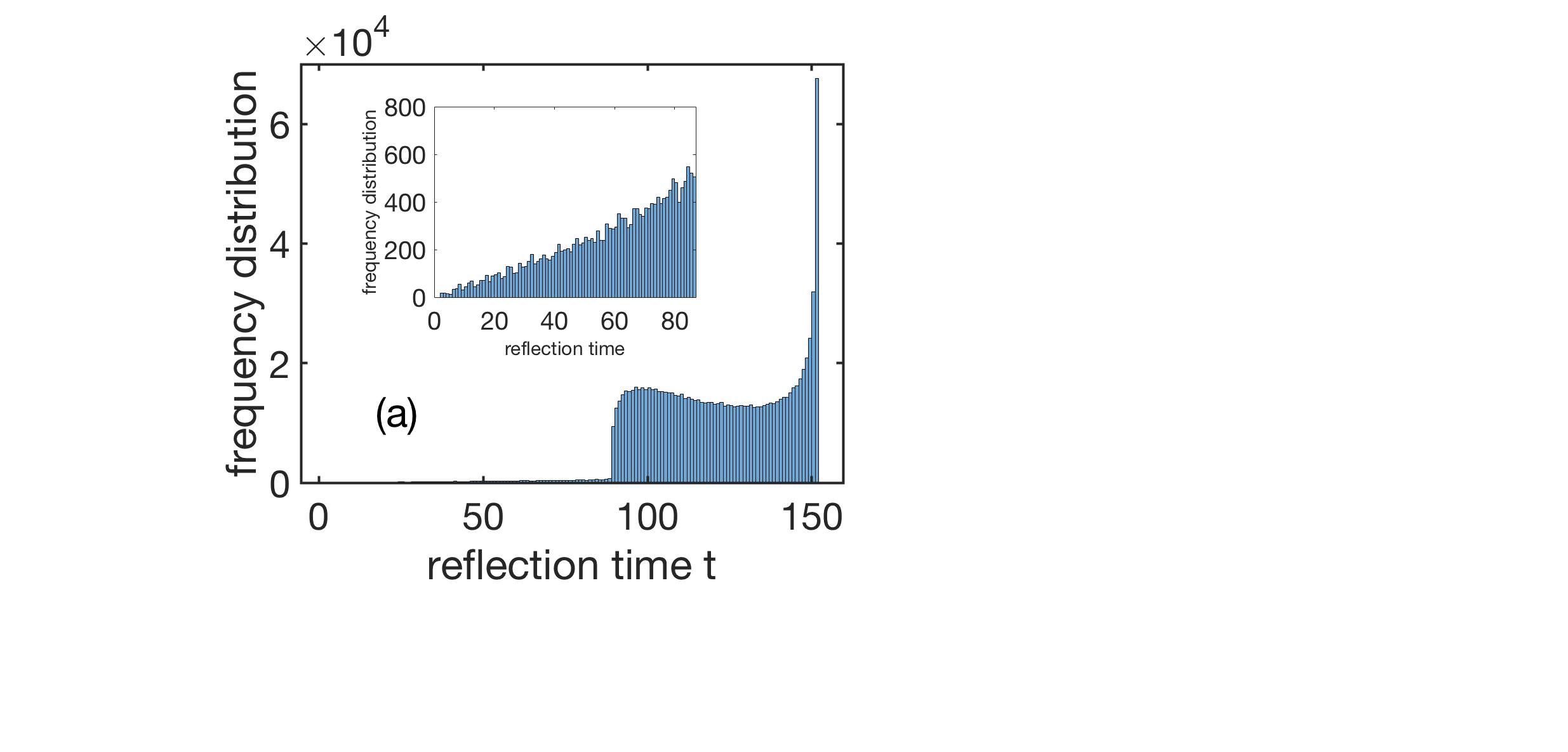}}
\hspace*{-1.6cm} \vspace*{-0.8cm} \parbox{6cm}{\vspace*{0.4cm} \includegraphics[width=14cm,height=8.0cm]{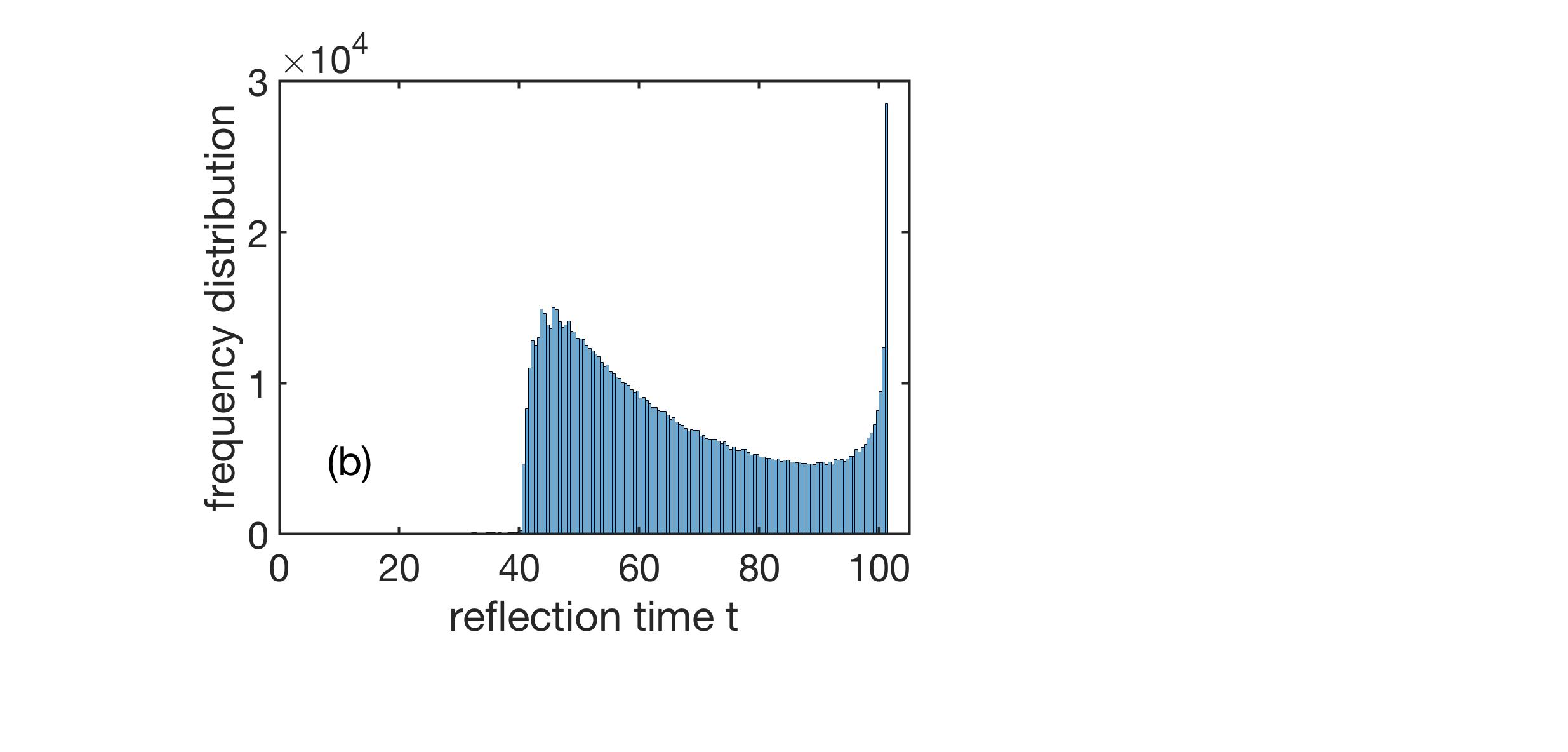}}\\
\parbox{6cm}{\vspace*{0.4cm} \includegraphics[width=13cm,height=7.0cm]{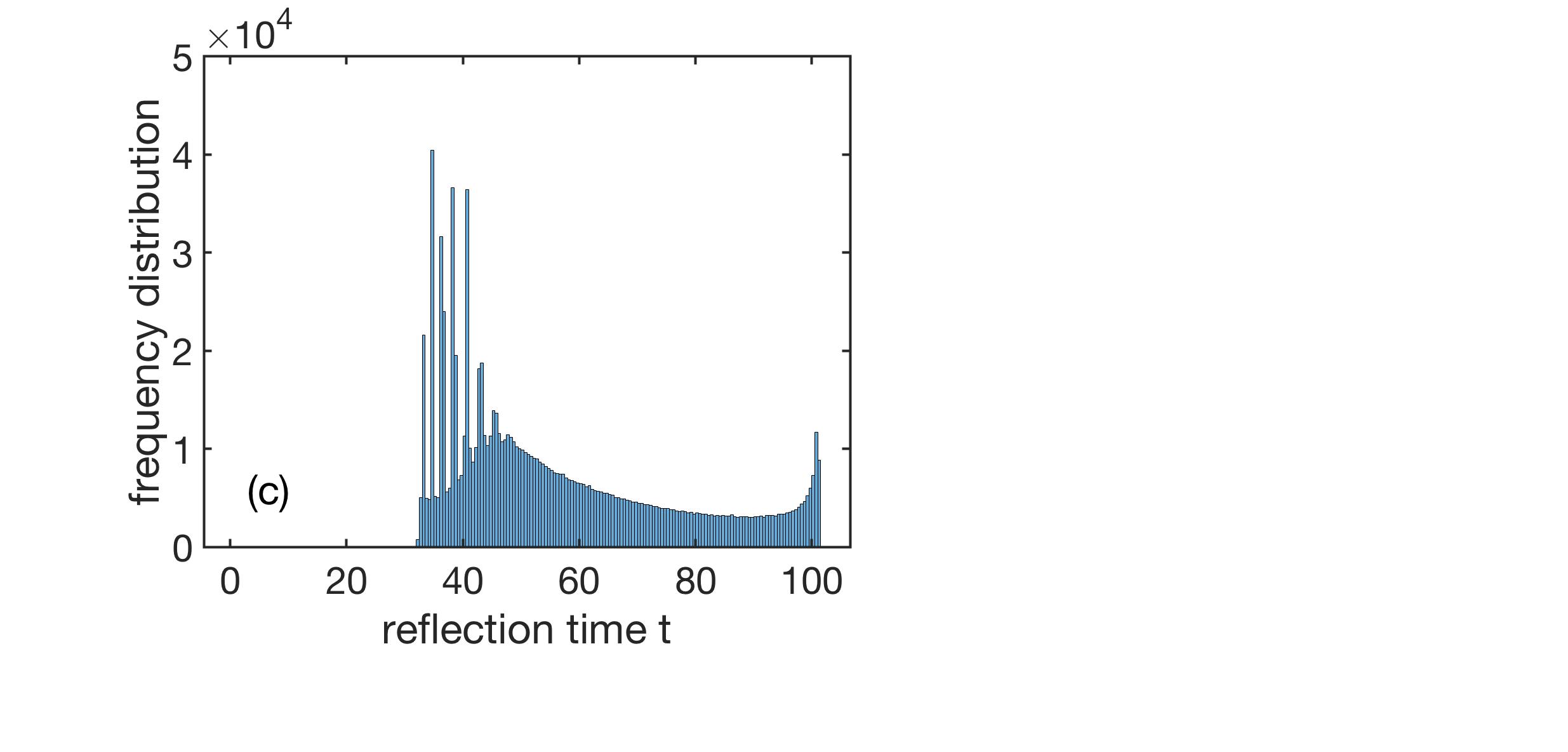}}
\caption{(a) Reflection time distribution for scattering in the confining channel (region I) 
for $E=0.1$. The ensemble of $10^6$ trajectories obeys the initial conditions $q_1=0,q_2=20$ and the kinetic
energies are chosen uniform randomly. (b,c) Same as (a) but for $E=0.5$ and $E=0.8$ respectively.}
\label{fig5}
\end{figure}

\begin{figure}
\hspace*{-3.8cm} \parbox{10cm}{\includegraphics[width=14cm,height=6.7cm]{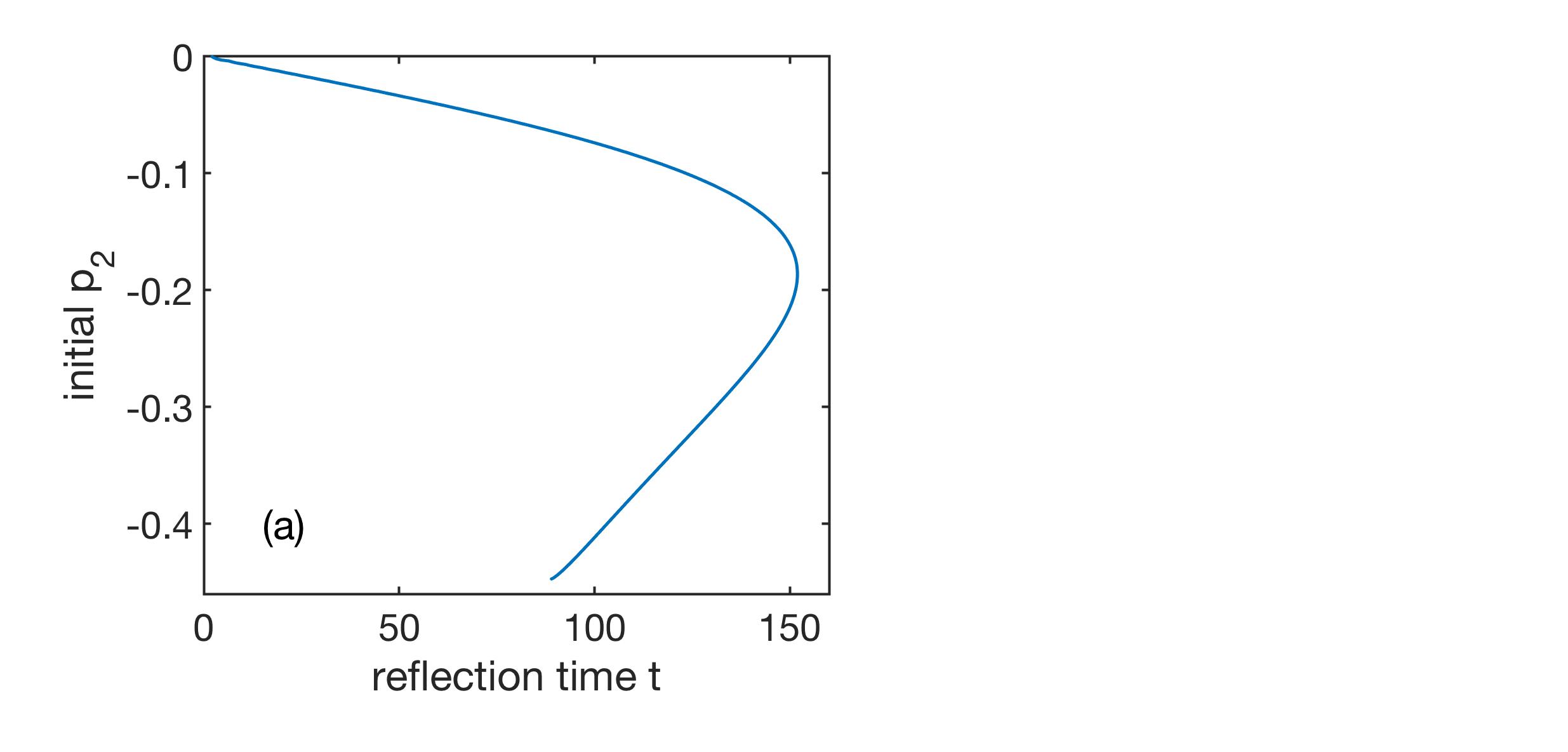}}
\hspace*{-1.6cm} \vspace*{-0.8cm} \parbox{6cm}{\vspace*{1.1cm} \includegraphics[width=14cm,height=8.0cm]{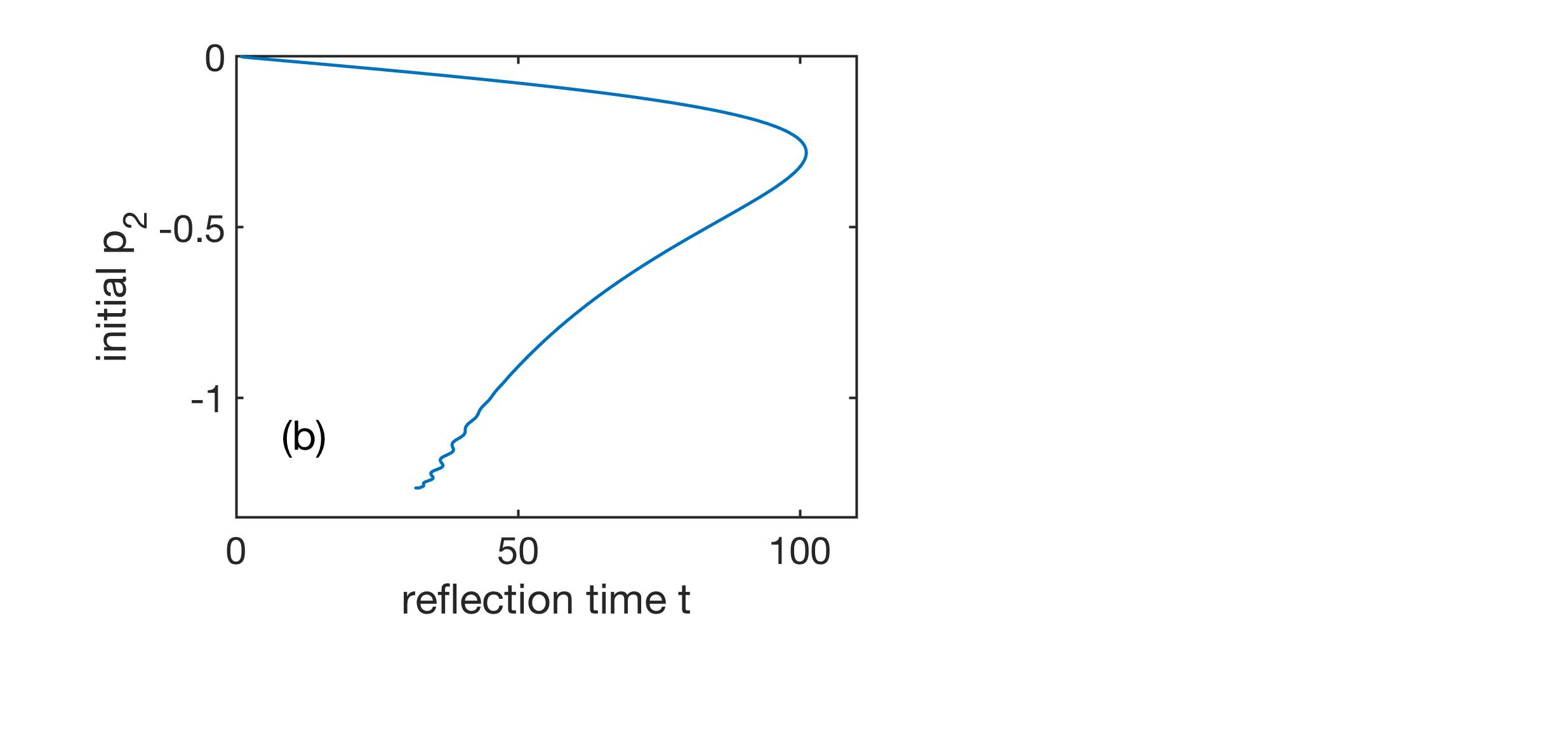}}\\
\caption{(a) Correlation diagram of the initial momentum $p_2$ versus the reflection time for $E=0.1$. (b) same
as in (a) but for $E=0.8$.}
\label{fig6}
\end{figure}

\subsection{Dynamics in the Confined Channel: Ensemble Properties} \label{sec:dynamicsb}

Lets explore the properties of ensembles of trajectories in order to gain a representative
view on the dynamics. In this subsection we focus on the confining channel of region I. We  
fix $q1,q2$ initially and random uniformly distribute the kinetic energy $E_{k2}$
(both the initially incoming and finally outgoing trajectory values for $q_2$ are fixed to 
$20$ which is sufficiently close to the asymptotic channel region 
where the transverse $q_1$ profile of the SEP is box-like).
$E_{k1}$ is then adapted to the energy shell and also uniform randomly distributed.
The random distribution of the corresponding momenta $p_1,p_2$ follows then a (piecewise) linear 
envelope. We proceed by first analyzing the case of low energies followed by higher energies in the 
confined channel.

We will proceed as follows in this subsection. First we will discuss the main features of the
reflection time distribution for the scattering in the confined channel. In a second step it will
be analyzed by employing so-called correlation diagrams. Subsequently we will explore the in-out
scattering functions followed by an analysis of the final kinetic energy distributions.

\subsubsection{Features of the reflection time distribution}

Figure \ref{fig5} shows the reflection time distribution (RTD) for an ensemble of $10^6$ trajectories. Inspecting
the case $E=0.1$ in Figure \ref{fig5}(a) our first observation is that the resulting frequency distribution
exhibits two plateaus and a dominant peak. For small reflection times $t < 90$ the RTD
is strongly suppressed (see inset of Figure \ref{fig5}(a)) and the distribution increases
linearly with an increasing reflection time. One reason for this
behaviour is the fact that the initial ($t=0$) momentum distribution of $p_2$ goes linearly to
zero at $p_2=0$ and therefore small forward momenta tentatively resulting in small reflection times 
are suppressed. At $t \approx 87$ (see Figure \ref{fig5}(a)) a sudden rise occurs for the RTD 
to a plateau of reflection times larger by more than an order of magnitude as compared to small 
reflection times. On this plateau the RTD shows a smooth oscillation with further increasing reflection
time that is stretched significantly towards larger reflection times. Then, for the maximum reflection time
$t \approx 155$ a dominant narrow peak is encountered. 

\subsubsection{Analysis via the correlation diagram}

What is the origin of the above features and in particular of the plateau structure of the RTD ?
To address this question, the correlation diagram of the reflection time depending on the
initial momentum $p_2 (t=0)$ is very instructive. Figure \ref{fig6}(a) shows this dependence for
the energy $E=0.1$. We observe that the 'mapping' of the initial momentum onto the reflection time
is represented by a well-defined curve. For small reflection times $t \lesssim 87$ this curve is single-valued
which is identical to the regime of the first plateau of reflection times of the RTD for low values.
From $t \approx 88$ on a double valuedness occurs i.e. there are two initial momentum branches that 
contribute to a certain interval of reflection times. At $t \approx 88$ the second (lower) branch starts at the
most negative possible value for the initial momentum $p_2(t=0)$ and moves to less negative values
with further increasing reflection time $t$. The latter momenta $p_2 \gtrsim -0.45$ possess a
very high probability in the initial uniform random ensemble and represent therefore a major reason 
for the above-mentioned step-like increase of the RTD. The point of the appearance of the second branch
in the ($p_2,t$) correlation diagram is the point of the appearance of trajectories that, starting
at $t=0$ from ($q_1=0,q_2=20$), travel all the way to the origin of the SEP (note that ${\cal{V}}(q_1=0,q_2)=0,
\forall q_2 > 0$). Since $E=E_{k2}$ holds in this case, i.e. all energy is kinetic energy of the 
degree of freedom $q_2$. The time for the occurence of this trajectory is straightforwardly determined
to be $t \approx 89$ which agrees with the above-observed value. The second lower branch of strongly negative
values of $p_2(t=0)$ represents a second dominant contribution to the reflection times beyond $t \approx 90$,
whereas the upper branch exists already for $t < 90$ and adds low probability to the RTD. 

The stretched oscillatory behaviour of the second plateau of the RTD in Figure \ref{fig6}(a) can be understood
by the different slopes of the two branches contributing to the statistics of the reflection time.
The lower branch possesses a steeper slope compared to the upper branch and its contribution to the
RTD thereby decreases more rapidly with increasing reflection time as compared to the corresponding impact
of the weaker slope of the upper branch. This leads on the second plateau to an overall decrease of the RTD.
The peak structure for even larger reflection times becomes here also understandable since the momentum
branch becomes tangentially vertical and therefore a broad range of momenta contribute to the maximal reflection
time. We note that the neighborhood of the onset of the lower branch in the correlation diagram ($p_2,t$) 
corresponds to strongly squeezed trajectories as discussed in subsection \ref{sec:dynamicsa}.

\begin{figure}
\hspace*{-3.8cm} \parbox{10cm}{\includegraphics[width=15cm,height=9.0cm]{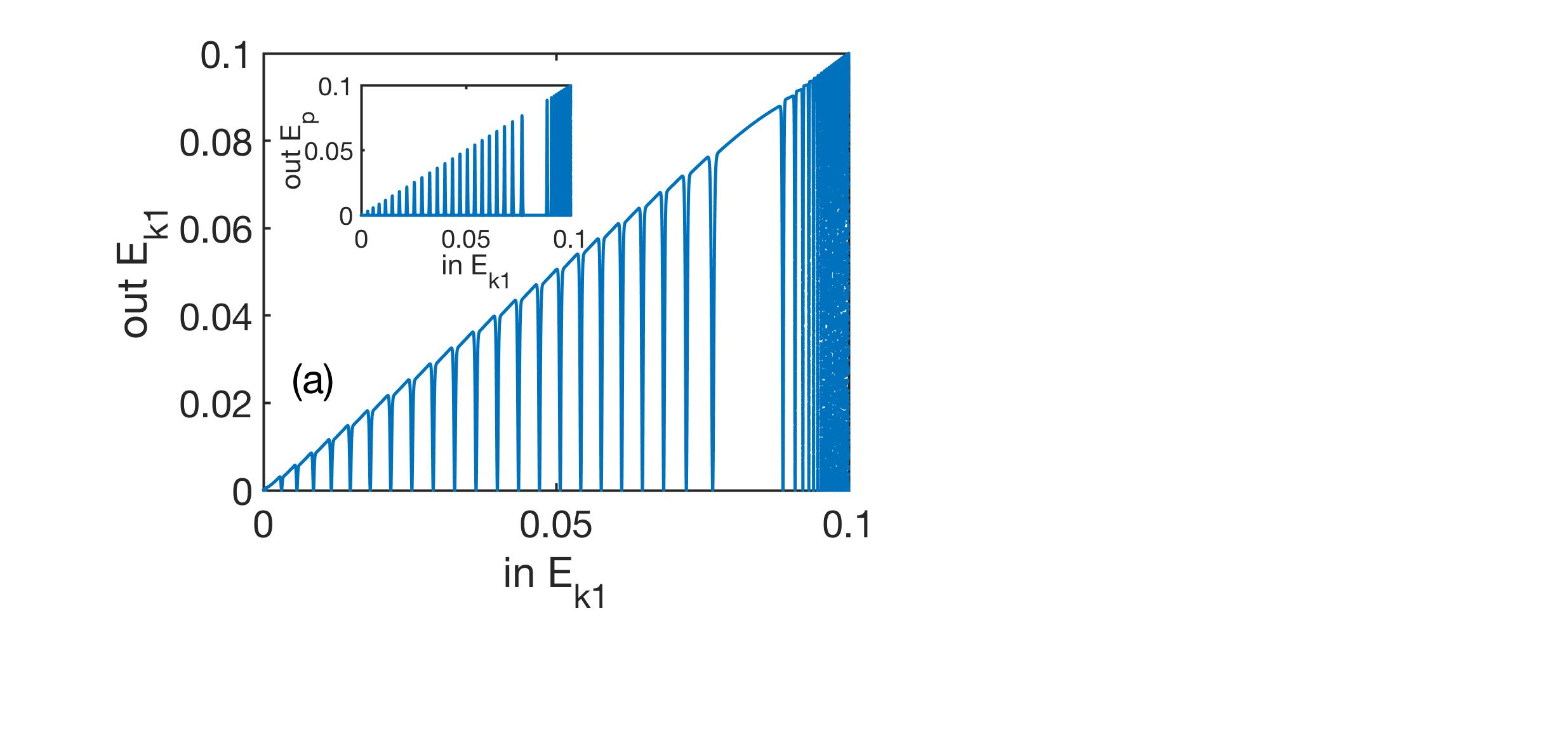}}
\hspace*{-1.8cm} \vspace*{-0.8cm} \parbox{6cm}{\vspace*{0.3cm} \includegraphics[width=14cm,height=8.0cm]{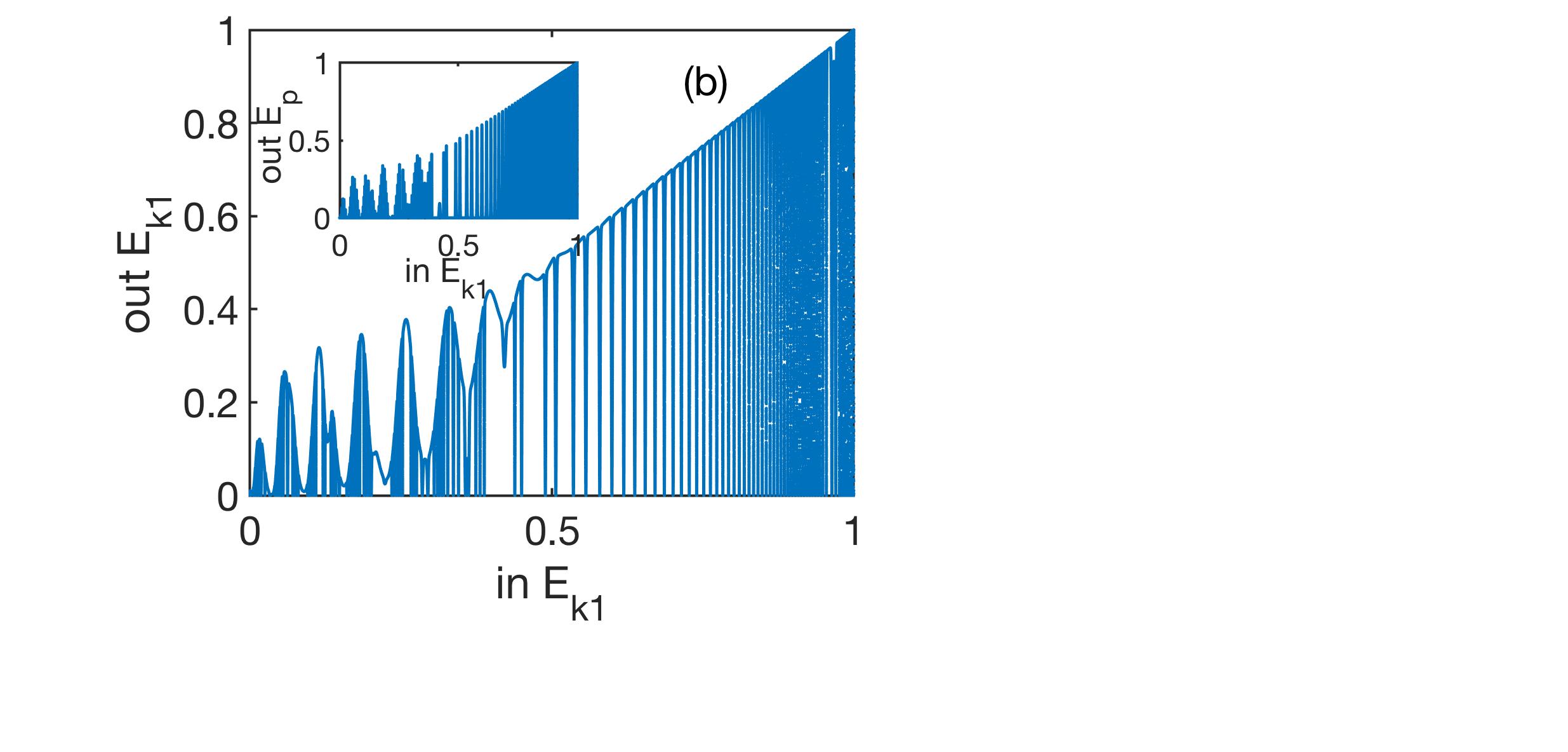}}\\
\caption{(a) In-out scattering function for the kinetic energy $E_{k1}$ for $E=0.1$. Inset:
Scattering function of incoming $E_{k1}$ and outgoing $E_p$. Initial conditions and ensemble properties
as indicated in Figure \ref{fig5}. (b) and its inset: Same as in (a) but for $E=1$.}
\label{fig7}
\end{figure}

\subsubsection{In-out scattering functions}

Sticking with the energy $E=0.1$ let us analyze some relevant in-out scattering functions related to the different kinetic and
potential energies. Figure \ref{fig7}(a) shows the scattering function, i.e. in-out mapping, of the kinetic energy $E_{k1}$.
Since at $t=0$ we have $E_p=0$ and $E= E_{k1} + E_{k2}$ the initial $E_{k1}$ directly translates to $E_{k2}$.
We observe that the outgoing $E_{k1}$ is proportional to the initial incoming $E_{k1}$ interrupted by narrow
dips or antipeaks. These antipeaks correspond to the situation where the outgoing potential energy becomes
maximal, see inset of Figure \ref{fig7}(a). This happens, due to the flatness of the bottom of the 
outgoing transversal potential well ($q_2=20$) and the corresponding steep walls, only if the phase of the
transversal $q_1$ oscillation of the outgoing trajectory is such that the particle coordinate $q_1$ encounters
those steep walls and converts its kinetic to potential energy. This process happens repeatedly when, in the above
sense, phase matching is encountered. Of course, the concrete value of the outgoing phase is determined 
also by the detailed scattering dynamics at small(er) values of $q_2$.

For large kinetic energies $E_{k1}$ first an energetic gap with no (anti-)peaks occurs and subsequently
an accumulating series of antipeaks for $E_{k1}$ and of peaks for $E_p$ is encountered (see Figure \ref{fig7}(a)).
This accumulation originates from the fact that the slow $q_2$-motion for large kinetic energies $E_{k1}$
and the resulting high frequency $q_1$ oscillations lead to an increasingly rapid change of the phase of the
outgoing trajectory with varying $E_{k2}$. These rapid phase changes lead also to a series of peaks with
large potential energies $E_p$ due to the repeated 'collisions' with the transverse potential walls for the
outgoing large $q_2$-value. The upper branch in the correlation diagram in Figure \ref{fig6}(a) shows for
small reflection times an approximately linear behaviour of $p_2$ as a function of the reflection time.
The corresponding kinetic energies $E_{k1},E_{k2}$ scale then quadratically with the reflection
time for sufficiently small initial momenta $p_2$. The accumulation of peaks at $E_{k1}=0.1$ (see Figure \ref{fig7}(a))
is connected to this quadratic scaling: varying $E_{k2} \propto t^2$ linearly leads to a corresponding
nonlinear phase change and accumulation of peaks. A final note is in order concerning the conversion
of the kinetic energy to potential energy in the course of the transversal channel dynamics. A brief calculation
shows that the ratio of the forces acting on the two degrees of freedom ($q_1,q_2$) reads as follows

\begin{equation}
\frac{\dot{p}_1}{\dot{p}_2} = \frac{q_2}{q_1 \ln |q_1|}
\end{equation}

which indicates that the force acting on $q_1$ dominates the force acting on $q_2$ at our incoming/outgoing boundary
since $q_2 >> |q_1|$. Therefore, the changes of the motion of $q_1$ are decisive for the conversion of the corresponding
kinetic energy to potential energy.

\begin{figure}
\hspace*{-3.8cm} \parbox{10cm}{\includegraphics[width=15cm,height=9.0cm]{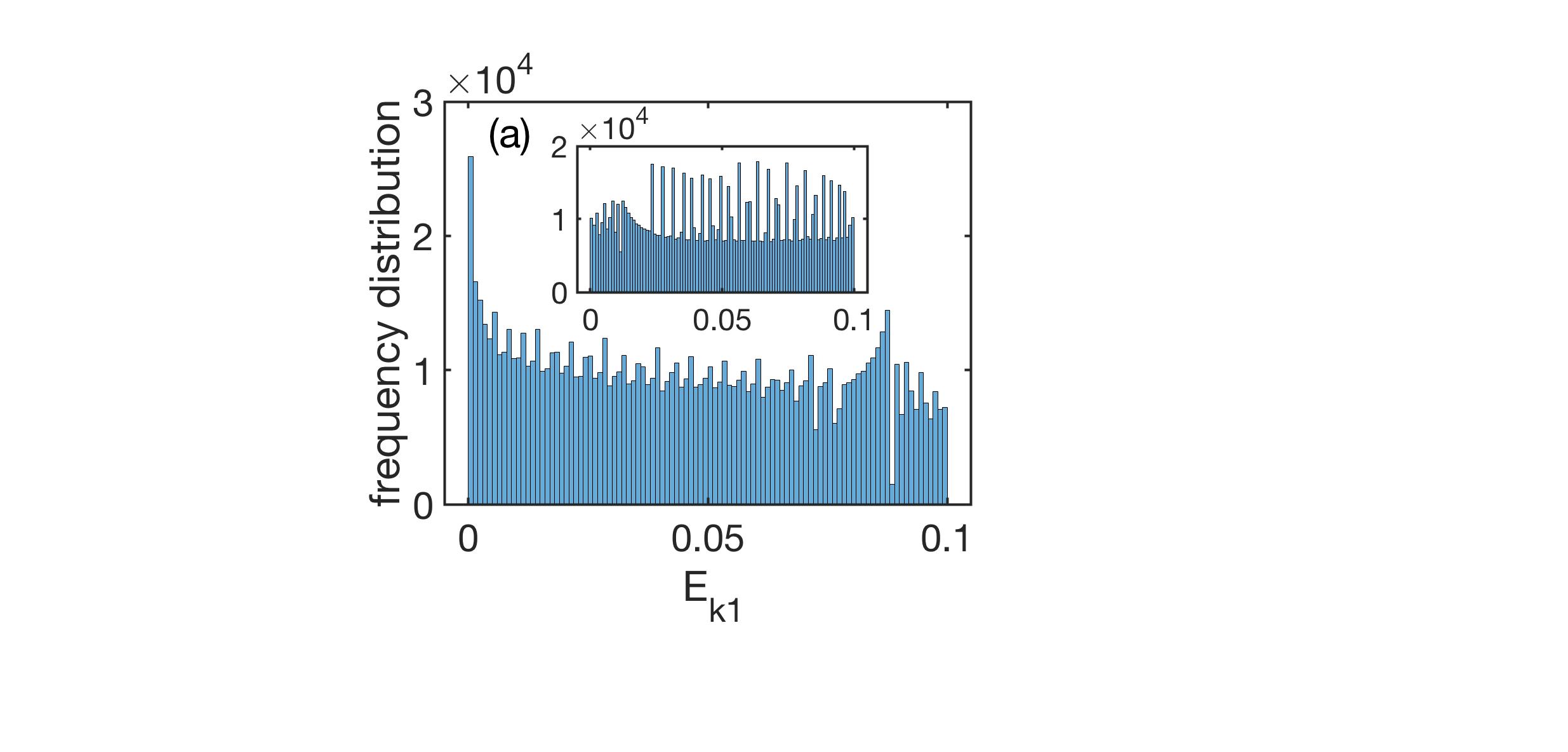}}
\hspace*{-0.3cm} \vspace*{-0.8cm} \parbox{6cm}{\vspace*{0.1cm} \includegraphics[width=14.5cm,height=8.5cm]{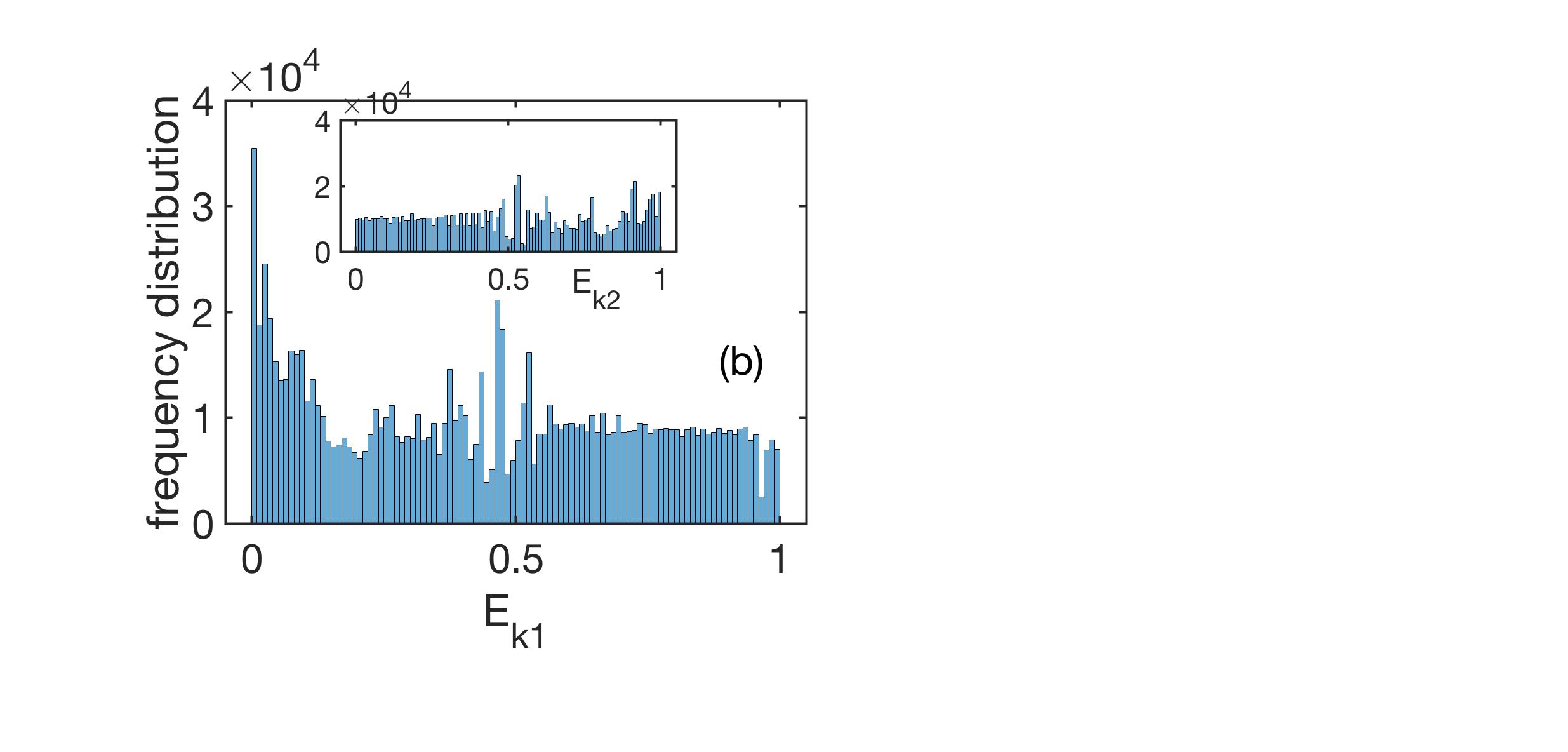}}
\caption{(a) Outgoing kinetic energy distribution of $E_{k1}$ for an ensemble of $10^6$ trajectories with the initial
conditions $q_1=0,q_2=20$ for $E=0.1$ and uniform randomly chosen initial kinetic energies $E_{k1},E_{k2}$. 
Inset: Corresponding outgoing kinetic energy distribution for $E_{k2}$. (b) Same as in (a) but for $E=1$.}
\label{fig8}
\end{figure}

\subsubsection{Kinetic energy distributions}

We explore now the outgoing kinetic energy distributions (KED) $E_{k1},E_{k2}$ of the scattering of the ensemble of
of trajectories in the CC in region I. Figure \ref{fig8}(a) shows these KED in the main figure and inset
for a total energy of $E=0.1$. Opposite to the uniform initial KED the scattered distributions exhibit 
several pronounced structural features. In particular we observe an oscillatory progression of fluctuations with
increasing kinetic energy $E_{k1}$ and correspondingly $E_{k2}$ (see inset). Narrow and broad peaks are located at the
maximal and minimal values of the kinetic energies, to a lesser extent at intermediate values. Besides high frequency
oscillatory structures also asymmetric peaks of one-sided smooth character appear, extending over a broader
interval of energies. These oscillations observed in the KED correspond to the oscillations observed in the
kinetic energy scattering functions discussed above (see Figure \ref{fig7}(a)). An enhanced probability of the 
KED occurs if the outgoing kinetic energy integrates over the rapidly varying antipeaks.

Let us now focus on the case of higher energies thereby addressing the most important changes as compared to
the above discussion for $E=0.1$. For the RTD in Figure \ref{fig5}(b) ($E=0.5$) we observe that the stretched
oscillation occuring for $E=0.1$ at the location of the second plateau develops into a deep valley while at the
same time the corresponding correlation diagram ($p_2(t=0),t$) deforms (not shown here for brevity). 
Moving on to the case $E=0.8$ the RTD (see Figure \ref{fig5}(c)) shows now at the point of the appearance
of the lower branch of the corresponding correlation diagram (see Figure \ref{fig6}(b)) a series of major
peaks which stem from the oscillatory behaviour at the beginning of this branch in the correlation diagram.
The latter can be understood by noticing that the contributions to the RTD stem either from straight
monotonic parts of the initial momentum $p_2$ (low values of reflection time probability) or from parts
around the turning point (high values of reflection time probability). The oscillations in the correlation
diagram therefore directly translate to the peaks of the RTD. In Figure \ref{fig7}(b) these modulations
are reflected in the in-out mapping of the kinetic energy distribution for $E_{k1}$ (for $E=1$). In
the corresponding inset of Figure \ref{fig7}(b) the modulations leave their fingerprints in the peak
structure of the in-out mapping of $E_{k1}$ to $E_p$. The outgoing KED for $E=1$ develops equally
a highly fluctuating behaviour.

\subsection{Dynamics Above the Saddle Points} \label{sec:dynamicsc}

We now investigate the dynamics for energies above the two saddle points, for which case both
reflection and transmission is possible. Our trajectories are initialized again in the CC in region
I of the SEP for comparatively large values of $q_2$ where the transverse profile of the channel
is already box-like. We will first inspect a few prototype trajectories and then focus on the
ensemble properties. The primary quantity of interest is the scattering distribution with respect
to the angle $\Phi = \arccos{\left(\frac{q_1}{q_1^2+q_2^2}\right)}$ and the fraction of reflected
versus left (region II) and right (region III) transmitted trajectories with varying energy.
Obviously, $\Theta$ characterizes the direction of scattering among the two degrees of freedom $q_1,q_2$.

\begin{figure}
\parbox{10cm}{\includegraphics[width=15cm,height=9.0cm]{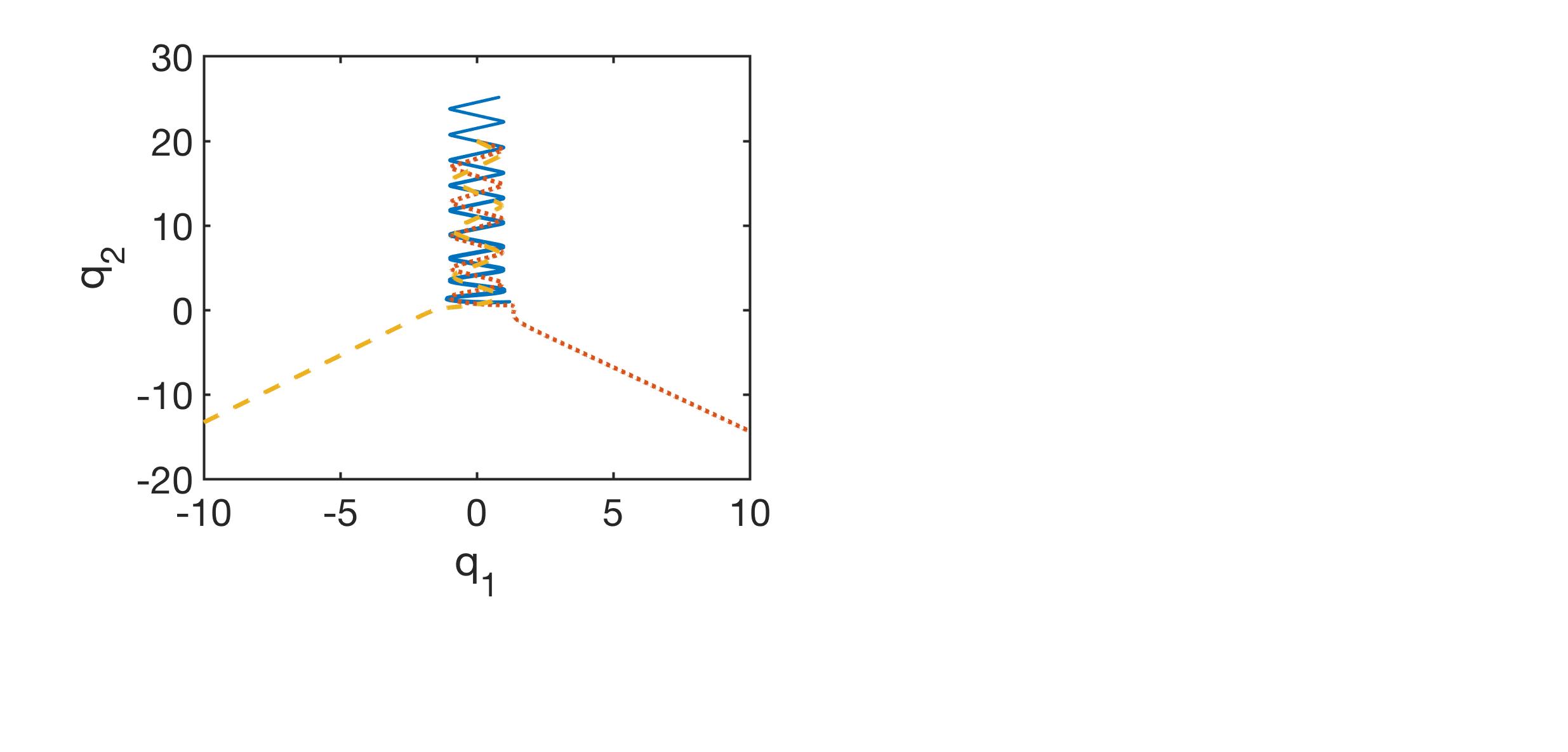}}\\ \vspace*{-2cm}
\caption{Three example trajectories in the ($q_1,q_2$) plane for initial conditions $q_1=0,q_2=20$ and
$p_2=-0.9,-1.1,-1.3$ for an energy $E=1.2$ above the saddle point energies.
The back reflection (region I) as well as the left (region II) and right (region III)
transmission processes are demonstrated.}
\label{fig9}
\end{figure}

Figure \ref{fig9} illustrates three prototypical trajectories in the ($q_1,q_2$) plane for an energy
$E=1.2$ above the saddle points. While one of them is backreflected into the CC, the other two are
transmitted, one of them to region II and the other one to region III. As indicated previously
(see section \ref{sec:hamiltonian}), region II ($q_1<0,q_2<0$) implies that the two particles move in a correlated
manner in the same direction whereas region III ($q_1>0,q_2<0$) is responsible for processes where both particles
move in opposite directions. The dynamics in the vicinity of the saddle points leads to the branching
to one of these regions. Back reflection means that the particle with coordinate $q_2$ leaves with
an asymptotically free motion to $-\infty$ whereas the second particle remains in an oscillating state.
Transmission yields then freely propagating particles asymtotically. In this sense a deconfinement
transition happens that turns the $q_1$ degree of freedom which is strongly confined in region I and strongly nonlinearly coupled
to the $q_2$ degree of freedom after passing via the saddle point to a free degree of freedom.
This means a dynamical unfolding of the confined degree of freedom $q_1$ while passing from
the channel region via the saddle points to the unconfined and asymptotically free region.
The two regions II and III are separated by a (singular) barrier centered around $q_1=0$ which
becomes an infinite square barrier asymptotically for $q_2 \rightarrow -\infty$.

\begin{figure}
\hspace*{-3.8cm} \parbox{10cm}{\includegraphics[width=13cm,height=8.3cm]{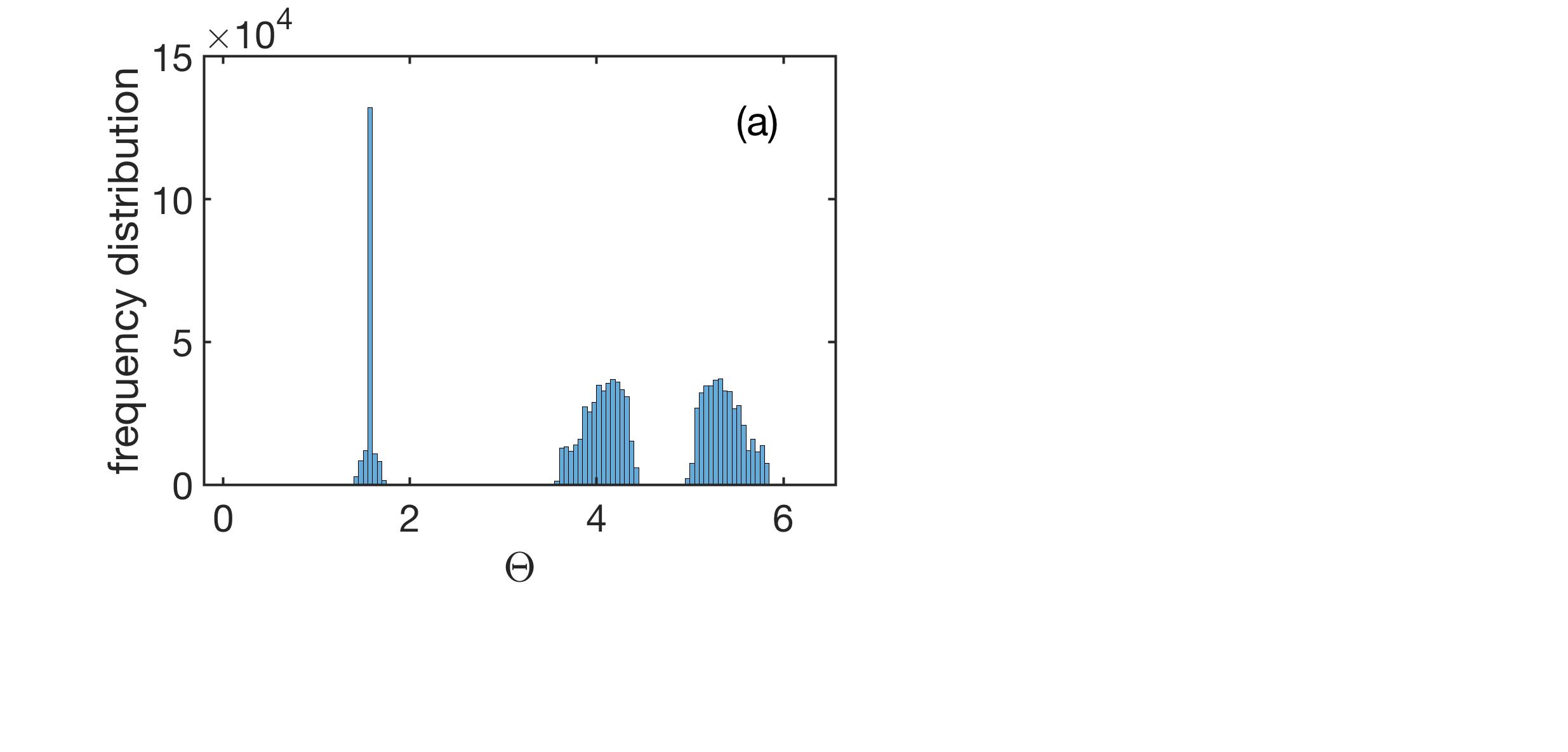}}
\hspace*{-3.1cm} \vspace*{-0.8cm} \parbox{6cm}{\vspace*{-0.1cm} \includegraphics[width=14.5cm,height=8.0cm]{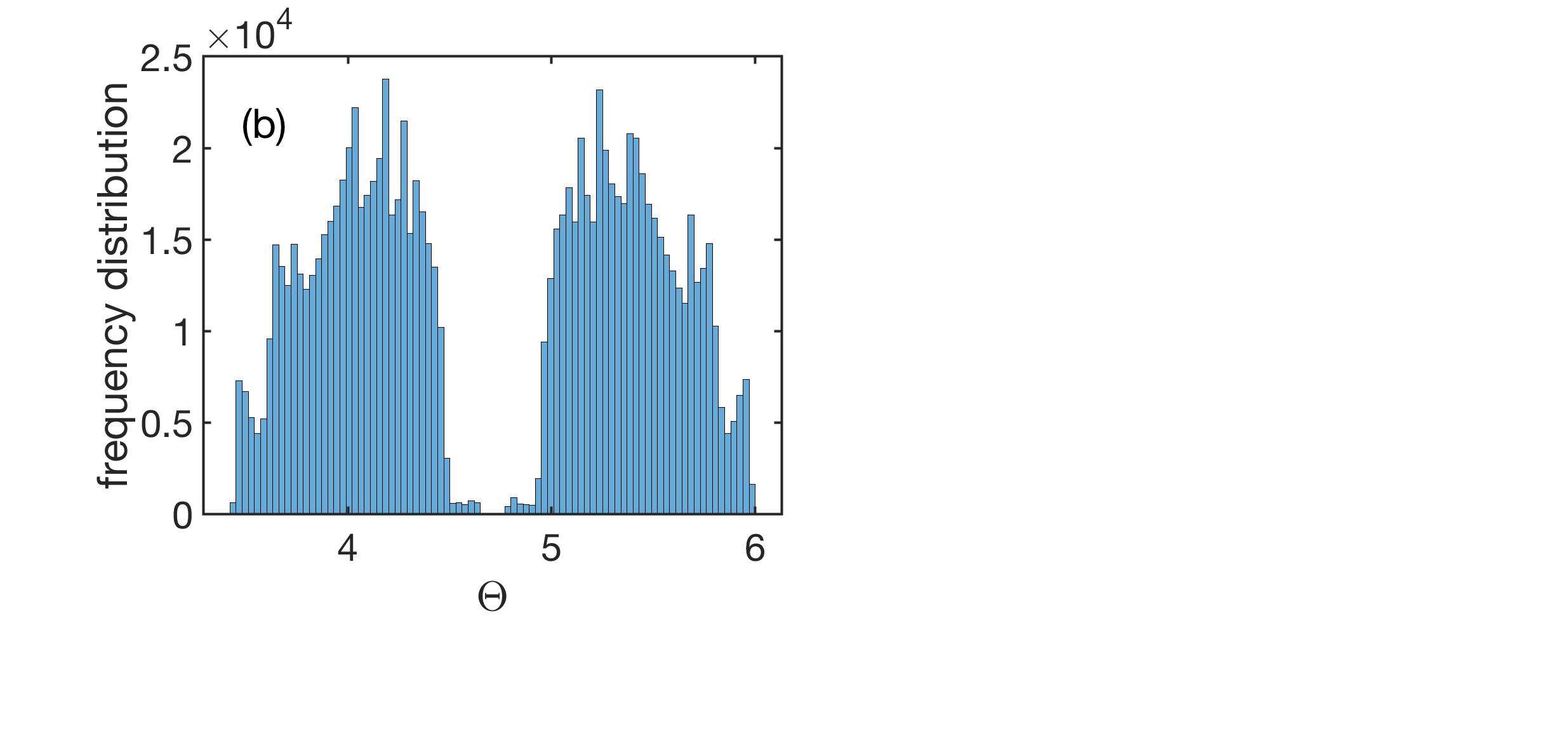}}\\
\vspace*{-0.5cm} \parbox{6cm}{\vspace*{-0.4cm} \includegraphics[width=14cm,height=8.0cm]{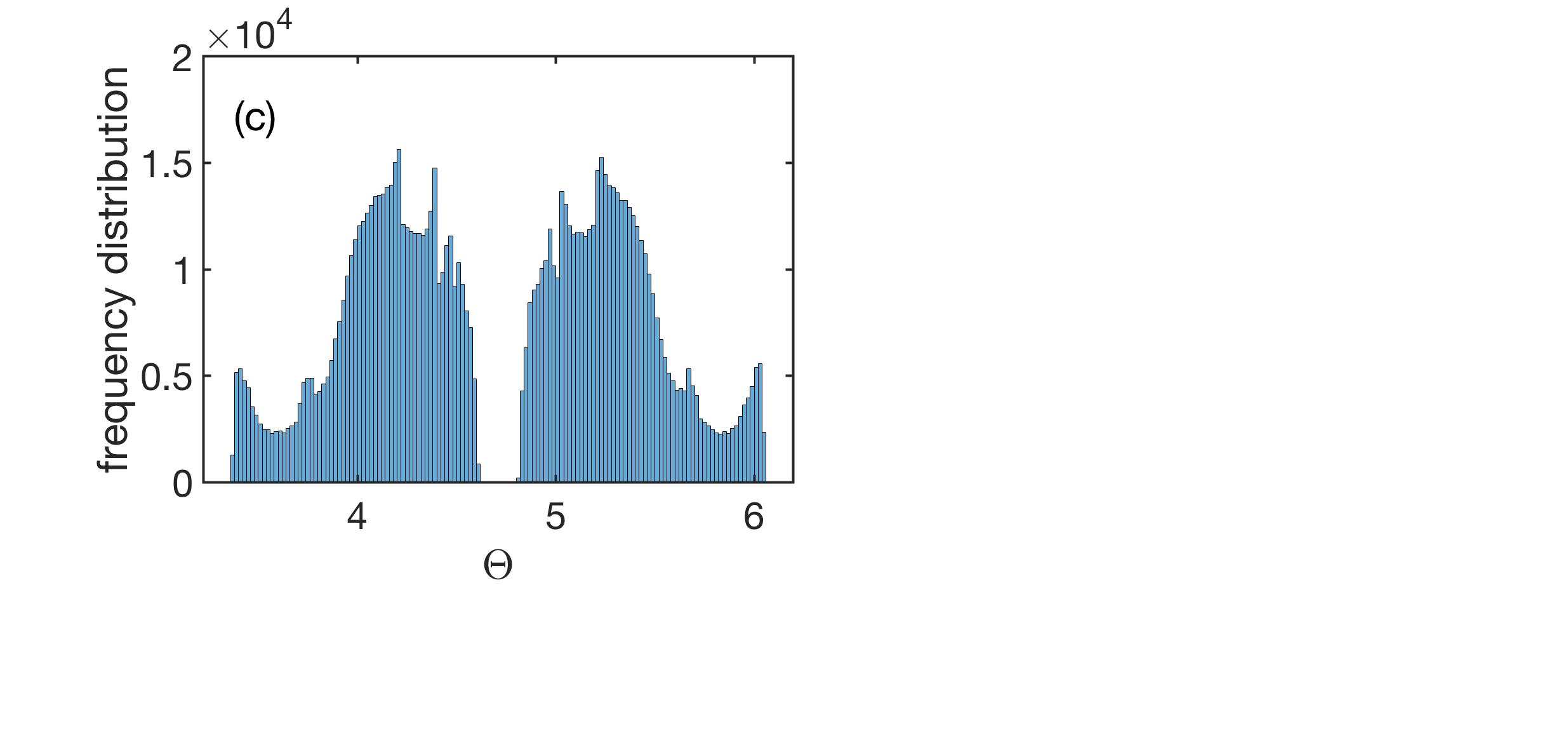}} \vspace*{-1.5cm}
\caption{Angular distribution functions of the scattering process for an ensemble of $10^6$ trajectories injected from the CC in
region I for the initial conditions $q_1=0, q_2=20$ for the energies $E=1.5,2.5,5.5$ in (a),(b),(c) correspondingly, which
are above the saddle point energies. $\Theta$ is provided in radian units.}
\label{fig10}
\end{figure}

Figure \ref{fig10}(a) shows the scattering angular distribution (SAD) for an energy $E=1.5$ slightly above the
saddle point energies. Three isolated peaks can be observed. The one located at $\frac{\pi}{2}$ is the dominant
and narrowest peak. It corresponds to trajectories backscattered into the CC. The other two smoother and broader
peaks correspond to the transmission scattering into the asymptotic regions II and III respectively. They are centered around
$1.3 \pi$ and $1.7 \pi$ respectively. Generally with increasing energy the height of the back scattering peak
decreases, while for the forward scattered distributions the corresponding widths increase and the shape becomes
increasingly asymmetric while the maximal values increase too. 

Focusing on the SAD in Figure \ref{fig10}(b) for $E=2.5$ the backscattering peak has disappeared completely
and two very broad distributions in the corresponding sectors of region II and III have emerged. They almost
overlap (note that there is an impenetrable barrier between them) and possess a characteristic asymmetric overall shape
imprinted by the presence and shape of the barrier: a steep slope occurs around $\Theta = \frac{3 \pi}{2}$ and a
smoother decay for the outer part of the distributions, though modulated by additional peaks. Finally, for even
higher energies (see Figure \ref{fig10}(c) for $E=5.5$) the distributions develop a dynamically induced modulation.

\section{Conclusions and Outlook} \label{sec:conout}

We have performed in this work a first step on the route to form complex structures from fundamental building
blocks with superexponential interactions by exploring a system of two degrees of freedom.
The underlying superexponential potential is of very simple appearance but shows an amazingly rich behaviour and
properties. Opposite to common two-body problems with e.g. Coulomb or dipolar interaction potentials 
depending on the relative coordinates of the two particles, the superexponential potential puts the nonlinearity
to the extreme: the base of its exponential dependence and the exponent depend on the dynamical degrees of
freedom. Resultingly the SEP is inherently inseparable, highly asymmetric and nonlinear and does not obey
the standard asymptotic boundary conditions. Specifically we have shown that the SEP exhibits three
different regions with a qualitatively different geometry and coupling of the degrees of freedom.
In the so-called region I we encounter a confined channel geometry along which the transversal confinement
continuously changes its anharmonicity from box-like in the asymptotics to an inverse cusp structure close 
to the origin. This channel is connected via two saddle points to the regions II and III which are separated
by a repulsive barrier. In the latter regions the dynamics is asymptotically free and corresponds to a 
correlated motion in the same or opposite directions of the two degrees of freedom respectively.
The scattering dynamics initialized in the confining channel therefore leads for energies above the saddle points
to a deconfinement transition and finally to asymptotic freedom. In this sense the originally confined
degree of freedom is dynamically unfolded within the highly nonlinear scattering process.

More specifically we have explored the scattering dynamics in the confining channel below and above the
saddle point energies for several relevant observables. The transition from a box-like to an extremely
squeezed channel towards the scattering center leads to a characteristic energy exchange pattern of series of
plateaus converging to a sequence of highly localized peaks. The reflection time statistics of the corresponding
ensembles show two major plateaus for low energies which can be analyzed and understood by employing
corresponding correlation diagrams between the initial momentum and the reflection time. For higher energies
additional peak structures occur. The dynamics above the saddle points connects the channel region with
the transmission regions and our analysis of the resulting angular distribution functions identifies
the characteristics of this scattering with varying energy. 

This work represents the basis for many possible extensions to come. Our interaction potential
${\cal{V}}(q_1,q_2)=|q_1|^{q_2}$ is nonreciprocal i.e. it doesnt treat the two degrees of 
freedom on an equal footing. A natural extension of the SEP would therefore be to symmetrize
it yielding ${\cal{V}}_s(q_1,q_2)=|q_1|^{q_2}+|q_2|^{q_1}$. This interaction potential 
possesses more than one channel and resultingly an even more intricate dynamics as compared
to the presently treated nonreciprocal case. Investigating this setup therefore goes beyond the
scope of this work. Our two degrees of freedom system represented by a single SEP term, can be extended in multiple ways to
several degrees of freedom. Indeed, the coupling between the exponent degrees of freedom and the
base degrees of freedom can be chosen in different ways, such that the analogue of artificial
'atoms', 'molecules', 'chains' or 'clusters' of networks of channels and regions of free motion being 
connected by saddle points could be imagined. Second the dependence on the absolute coordinates
seems to be very natural, but not the only possibility: relative coordinate dependencies might
introduce a qualitatively different behaviour. The generalization to higher dimensions opens the
route to exploit radial and angular coordinate dependencies in order to control possible anisotropies.

Finally the question can be posed what physical systems can potentially be described by our superexponential
interaction profile. While we do not have a conclusive answer to this question it is conceivable that
the SEP is a result of an effective description of an already complex system consisting of many
cooperative degrees of freedom. On the other hand the dynamical degrees of freedom appearing in the
SEP could also describe the motion in some parameter or effective space and not the real coordinate space.

\section{Acknowledgments}

This work has been in part performed during a visit to the Institute for Theoretical Atomic, Molecular and Optical Physics (ITAMP)
at the Harvard Smithsonian Center for Astrophysics in Cambridge, Boston, whose hospitality is gratefully acknowledged.
The author thanks F.K. Diakonos and B. Liebchen for a careful reading of the manuscript and valuable comments.


\begin{thebibliography}{99}
\bibitem{Friedrich} H. Friedrich, Theoretical Atomic Physics, Springer International Publishing 4th ed. 2017.
\bibitem{Helgaker} T. Helgaker, P. Jorgensen and J. Olsen, Molecular Electronic Structure Theory, John Wiley
and Sons, 2nd ed. 2020.
\bibitem{Jellinek} J. Jellinek, Theory of Atomic and Molecular Clusters, Springer New York 1999.
\bibitem{Natelson} D. Natelson, Nanostructures and Nanotechnology, Cambridge University Press, Cambridge 2015.
\bibitem{Ashcroft} N.W. Ashcroft and N. Mermin, Solid State Physics, Brooks Cole Publishing Company, Singapure, 2018.
\bibitem{Wilson} E.B. Wilson, J.C. Decius, P.C. Cross, Molecular Vibrations: The Theory of Infrared and Raman Vibrational
Spectra, Dover Books on Chemistry, 1980.
\bibitem{Tabor} M. Tabor, Chaos and Integrability in Nonlinear Dynamics, John Wiley and Sons (1989).
\bibitem{Strogatz} S.H. Strogatz, Nonlinear Dynamics and Chaos: With Applications to Physics, Biology, Chemistry,
and Engineering (Studies in Nonlinearity), Westview Press 2nd Edition (2015).
\bibitem{Reichl} L.E. Reichl, The Transition to Chaos, 2nd edition Springer (2004).
\bibitem{Flach} S. Flach, C.R. Willis, Phys.Rep. 295, 181 (1998).
\bibitem{Cao} L. S. Cao, D. X. Qi, R. W. Peng, Mu Wang, and P. Schmelcher, Phys.Rev.Lett. 112, 075505 (2014).
\bibitem{Schmelcher1} P. Schmelcher, Phys. Rev. E 98, 022222 (2018).
\bibitem{Schmelcher2} P. Schmelcher, arXiv:1909.09792, acc.f.publ. J.Phys.A
\end{thebibliography}
\end{document}